\documentclass[11pt]{article}
\usepackage[nosort]{cite}
\usepackage{dsfont}
\usepackage{amsfonts}
\usepackage{amssymb}
\usepackage{amsmath}
\usepackage{graphicx}
\usepackage{epstopdf}
\usepackage{slashed}
\usepackage{setspace}
\usepackage{multirow}

\newcommand{\compactsubsection}[1]{\vspace{2mm}\noindent {\bf #1}\\[1mm]}

\usepackage{color}

\newcommand{\cn}{{\cal N}}

\def \t{\tilde t}

\newcommand{\reef}[1]{(\ref{#1})}

\newcommand{\labell}[1]{\label{#1}}

\newcommand{\mt}[1]{\textrm{\tiny #1}}

\newcommand{\eg}{{\it e.g.,}\ }
\newcommand{\ie}{{\it i.e.,}\ }

\newcommand{\suv}{{$_\mt{UV}$}}
\newcommand{\sir}{{$_\mt{IR}$}}

\newcommand{\hg}{\widehat{g}}
\newcommand{\hR}{\widehat{R}}

\newcommand{\vphi}{\varphi}

\newcommand{\dash}{\text{-}}

\newcommand{\ca}{{\cal A}}

\newcommand{\be}{\begin{equation}}
\newcommand{\ee}{\end{equation}}
\newcommand{\nn}{\nonumber}

\def\be{\begin{equation}}
\def\ee{\end{equation}}
\def\bea{\begin{eqnarray}}
\def\eea{\end{eqnarray}}
\def\ba{\begin{array}}
\def\ea{\end{array}}
\def\bd{\begin{displaymath}}
\def\ed{\end{displaymath}}

\def\a{\alpha}
\def\b{\beta}

\def\d{\delta}
\def\e{\epsilon}           
\def\vf{\varphi}  
\def\g{\gamma}
\def\h{\eta}

\def\k{\kappa}                    
\def\l{\lambda}
\def\m{\mu}
\def\n{\nu}
\def\o{\omega}  
\def\r{\rho}                                     
\def\s{\sigma}                                   
\def\t{\tau}
\def\u{\upsilon}
\def\x{\xi}

\def\D{\Delta}

\def\G{\Gamma}

\def\O{\Omega}

\def\pa{\partial}                              
\def\>{\rangle} 
\def\<{\langle} 
\def\Dsl{D \hskip-.6em \raise1pt\hbox{$ / $ } }
\def\to{\rightarrow}
\def\pa{\partial}

\def\lab{\label}

\newcommand{\eps}{\epsilon}
\newcommand{\lra}{\leftrightarrow}

\begin{document}

\setstretch{1.05}

\begin{titlepage}

\begin{flushright}
MCTP-12-09\\
MIT-CTP-4362  \\
PUPT-2413\\
SU-ITP-12/14
\end{flushright}
\vspace{.1cm}

\begin{center}
\bf \Large
On renormalization group flows and the $a$-theorem in 6d
\end{center}
\vspace{2mm}
\begin{center}
{\bf Henriette Elvang${}^{a}$,
Daniel Z.~Freedman${}^{b,c,d}$, Ling-Yan Hung${}^{e}$,} \\[1mm]
{\bf Michael Kiermaier$^{f}$, Robert C.~Myers${}^{e}$, Stefan Theisen${}^{g}$} \\
\vspace{0.7cm}
{${}^{a}${\it Randall Laboratory of Physics, Department of Physics,}\\
{\it University of Michigan, Ann Arbor, MI 48109, USA}}\\[2mm]
{${}^{b}${\it Department of Mathematics, }{${}^{c}${\it Center for Theoretical Physics,}}\\
{\it Massachusetts Institute of Technology, Cambridge, MA 02139, USA}}\\[2mm]
{${}^{d}${\it   Stanford Institute for Theoretical Physics, Department of Physics,\\ Stanford University, Stanford, CA 94305, USA}}
\\[2mm]
{${}^{e}${\it Perimeter Institute for Theoretical Physics, Waterloo, Ontario N2L 2Y5, Canada}}
\\[2mm]
{${}^{f}${\it Joseph Henry Laboratories, Princeton University, Princeton, NJ 08544, USA}}
\\[2mm]
{${}^{g}${\it Max-Planck-Institut f\"ur Gravitationsphysik, Albert-Einstein-Institut, 14476 Golm, Germany}}\\[4mm]
{\small \tt  elvang@umich.edu,
 dzf@math.mit.edu, jhung@perimeterinstitute.ca, mkiermai@princeton.edu,
 rmyers@perimeterinstitute.ca, stefan.theisen@aei.mpg.de}
\end{center}
\vspace{-2mm}

\begin{abstract}
We study the extension of the approach to the $a$-theorem of Komargodski
and Schwimmer to quantum field theories in $d=6$ spacetime dimensions. The
dilaton effective action is obtained up to 6th order in derivatives. The
anomaly flow $a_{\rm UV} - a_{\rm IR}$ is the coefficient of the 6-derivative
Euler anomaly term in this action. It then appears at order $p^6$ in the low energy limit of  $n$-point scattering amplitudes
of the dilaton for $n\ge4$. The detailed structure with the correct anomaly coefficient is confirmed by direct calculation in two examples:
(i)  the case of explicitly broken conformal symmetry is illustrated  by the
free massive scalar field, and
(ii) the case of spontaneously broken conformal symmetry is demonstrated by the (2,0) theory on the Coulomb branch. In the latter example, the dilaton is a dynamical field so 4-derivative terms in the action also affect  $n$-point amplitudes at order $p^6$. The calculation in the (2,0) theory is done by analyzing an M5-brane probe in  AdS$_7 \times S^4$.

Given the confirmation in two distinct models, we attempt to use dispersion relations to prove that the anomaly flow is positive in general.
Unfortunately the 4-point matrix element of the Euler anomaly is
proportional to $stu$ and vanishes for forward scattering. Thus the optical
theorem  cannot be applied to show positivity. Instead  the anomaly flow is given
by a dispersion sum rule in which the integrand does not have definite
sign. It may be possible to base a proof of the $a$-theorem on the
analyticity and unitarity properties of the 6-point function, but our
preliminary study reveals some difficulties.
\end{abstract}

\end{titlepage}

\vspace{4mm}
\setstretch{0.3}
\setcounter{tocdepth}{2}
\tableofcontents
\setstretch{1.05}

\newpage

\setcounter{equation}{0}
\section{Introduction}

There is a common paradigm in quantum field theory in which the correlation functions of a non-conformal theory approach those of a conformal  theory, the CFT$_\text{UV}$,  at short
distance and those of another conformal theory, the CFT$_\text{IR}$, at long
distance.   The two CFT's are viewed as  end-points of the renormalization group flow (RG flow). Among the quantities characterizing a CFT are its trace anomaly coefficients,  frequently called central charges,
obtained by embedding the theory in a curved background metric $g_{\m\n}(x)$.

Zamolodchikov's $c$-theorem \cite{zamo} revealed a remarkable structure
of RG flows of two-dimensional
quantum field theories.  He showed that there is a positive definite function $C(g_i)$ on the space of couplings  that satisfies three properties: (a) $C$ decreases
monotonically along RG flows, (b) fixed points of the RG flow are
critical points for $C$, \ie $\partial_{g_i}C|_{g^*_j}=0$, and (c) at
these fixed points, $C(g_i^*)$ coincides with the central charge $c$ of
the corresponding conformal field theory. These properties hold in any
unitary, renormalizable and Lorentz invariant two-dimensional QFT.
A direct consequence is that for any RG flow, the central charges of the end-point CFT's satisfy the inequality
 \be
c_\mt{UV}\, \geq\, c_\mt{IR}\,.
 \labell{introx1}
 \ee
The central charge is often interpreted as providing
a measure of the `number of degrees of freedom' and hence the c-theorem
confirms the intuition that this number should  decrease along
RG flows.

Zamolodchikov's result motivated the search for a similar property in higher-dimensional quantum field theory.  For $d=4$ the trace anomaly
has two central charges $c$ and $a$ and takes the form
\be
\langle\,
T^\mu{}_\mu\,\rangle
~=~
c\,W^2 -
a\, E_4
\,,
 \labell{trace4}
 \ee
where $W^2$ denotes the square of the Weyl tensor  $W_{\mu\nu\rho\sigma}$ of the background geometry and $a$ multiplies the quadratic combination of curvatures which is the integrand of the Euler integral invariant
\be
\label{E4}
E_4
~=~
R_{\m\n\r\s}R^{\m\n\r\s}-4R_{\m\n}R^{\m\n}+R^2 \, .
\ee
In 1988 Cardy \cite{cardy} conjectured, with evidence from several models,  that for any RG flow,
it is the Euler central charge $a$ and its analogue in higher even spacetime dimension that satisfies the desired inequality
 \be
a_\mt{UV}\, \geq\, a_\mt{IR}\,.
 \labell{introx2}
 \ee

For several reasons it was  a difficult problem to prove this conjecture, although the
unsuccessful 25 year effort has taught us quite a bit of interesting
physics.  For example,  the central charges of many supersymmetric gauge theories in four dimensions can be calculated
\cite{anselmi2,anselmi1,ken}, and the Euler central charge $a$
 satisfies \reef{introx2}. Furthermore, it is quite easy to prove the $a$-theorem for QFT's which have an AdS/CFT dual \cite{flow2,gubser,cthem,jtl}.

A concise and insightful proof for any
four-dimensional RG flow connecting two conformal fixed points was
recently presented by Komargodski and Schwimmer in \cite{zohar} and discussed further in \cite{Z2,joep2}.
 The key idea in this work is to  use the dilaton field $\t(x)$ to probe the trace anomaly.  If conformal symmetry is spontaneously broken, as on the Coulomb branch of $\cn=4$ SYM theory,
the dilaton is a massless Goldstone boson in the spectrum of the theory.
If conformal symmetry is broken explicitly by dimensionful parameters of the QFT, the dilaton is introduced as a conformal compensator. As we review in Sec.~2 below, the $a$-theorem can be proved quite elegantly in this framework: it follows from the analyticity and unitarity of the forward 4-point scattering amplitude of the dilaton.

It is  natural to ask whether the dilatonic approach of \cite{zohar} can be extended to  QFT's in any even spacetime dimension. For general $d=2n$, the trace anomaly
can be written as \cite{traca,deser}\footnote{Scheme dependent terms of the form
$\pa_\m J^\m$ are ignored in \reef{trace4} and \reef{trace}. See
\cite{anselmi4,anselmi3} for an interesting application.}
 \be
\langle\,T^\mu{}_\mu\,\rangle = \sum_i c_i\,I_i -\,(-)^{d/2}a\, E_d
\,,
 \labell{trace}
 \ee
and it defines a number of  central charges of the CFT.
Each term on the right-hand side is constructed from
the background geometry and has conformal weight $d$. In particular,
$E_d$ is the Euler density in $d$ dimensions while $\sqrt{-g}I_i$ are conformal invariants. It is the Euler central charge which Cardy  \cite{cardy} conjectured to satisfy an $a$-theorem. The question that
originally motivated the present paper was whether the new insights
of \cite{zohar} can be applied to prove such a  theorem
for RG flows in $d=6$ dimensions.

We begin our investigation by constructing the $d=6$ effective dilaton action in Sec.~\ref{s:dilact}. As in $d=4$, the anomaly flow $\D a= a_\text{UV} -a_\text{IR}$  appears as the coefficient of the dilaton Wess-Zumino term which arises from the 6-derivative Euler density.\footnote{In this paper we assume that RG fixed point theories are conformal.  A scenario in which fixed point theories are scale invariant but not conformal invariant has been discussed in the recent literature, see \cite{scale,jeff} (and see also \cite{joep}), 
but are argued to be ruled out in $d=4$  \cite{joep2}.}
 A new feature appears in $d=6$: there are Weyl-invariant 4-derivative terms that  contribute to the dilaton scattering amplitudes. 
 We present the general 4-, 5- and 6-point on-shell dilaton scattering amplitudes in the low-energy expansion, specifically focusing on the contributions of order $p^4$ and $p^6$ in the momenta.

The structure of the dilaton effective action and scattering amplitudes is confirmed in explicit examples. In Sec.~\ref{s:massive} we study a free massive scalar and verify that when the massive mode is integrated out, the low-energy effective action of the dilaton (which is introduced as a conformal compensator in this case) agrees with our general result.

Encouraged by the massive scalar example, we discuss in Sec.~\ref{s:dispersiona} the structure of dispersion relations for the dilaton amplitudes. Unfortunately, no positivity statement or monotonic $a$-function can be extracted from the 4-point amplitudes at order $O(p^6)$.  It may be possible
to derive an $a$-theorem for RG flows in 6d from the 6-point dilaton amplitude,  but its analyticity, crossing, and unitarity properties are quite complicated, \eg see \cite{russians}, and we leave such an analysis for the future.

It is difficult to find conventional interacting field theories with tractable
RG flows
in $d=6$ dimensions. In $d=4$, the prime example for a theory with spontaneously broken conformal symmetry is $\mathcal{N}=4$ SYM on the Coulomb branch. The analogue  in $d=6$ is the M5-brane (2,0) theory on the Coulomb branch. It has no  weakly coupled description and is not easy to treat directly. For this reason, we turn to a holographic description. This system is holographic sine qua non, but
we will be testing the dilatonic formulation of the $a$-theorem in a
different way from traditional holographic $a$-theorems
\cite{gubser,flow2,cthem}. As a warm-up to the 6d (2,0) theory, we present in Sec.~\ref{s:DBI4d} a detailed account of how the 4d dilaton effective action is
extracted from the holographic treatment of the Coulomb branch RG flow of  $\mathcal{N}=4$ SYM.\footnote {We thank Juan Maldacena  for this suggestion.}  The techniques are then extended to the Coulomb branch flow for the  6d (2,0) theory in Sec.~\ref{s:DBI6d}.  Recently, ref.~\cite{sethi} studied how supersymmetry determines $\Delta a \sim N^2$ for the 6d $(2,0)$ theory.  
In our analysis, we match the full dilaton effective action from the DBI action of a probe M5-brane. In particular, we extract  $\Delta a$ at large $N$  including the numerical coefficient and find that it agrees with the literature \cite{baste}. Our analysis also clarifies the role of the dilaton in the case of spontaneously broken conformal symmetry: the dilaton must then be treated as a dynamical field, as opposed to the scenario of explicitly broken conformal symmetry in which the dilaton may be treated as a source \cite{Z2,joep2}. The difference is manifest in the dilaton amplitudes.

We conclude with a brief  summary  of our results and future directions in Sec.~\ref{s:discuss}.  Technical details are relegated to appendices:  App.~\ref{useful} collects conventions and some exact forms of 6d Weyl-invariants. App.~\ref{app:Seuler} contains two complementary derivations of the 6d Euler anomaly action. And finally App.~\ref{app:DBI2d} summarizes the Coulomb branch flow in the D1-D5 system at large $N$; this is the 2d analogue of our analyses of the 4d $\mathcal{N}=4$ SYM and the 6d (2,0) theory.


\setcounter{equation}{0}
\section{The $a$-theorem for $d=4$}
\label{s:D4}

We review the proof of the $a$-theorem in four dimensions presented by Komargodski and Schwimmer \cite{zohar}.
Issues relevant to the analysis in higher dimensions will be highlighted.
The key ingredient to understand is the role of the
`dilaton'.
This is conceptually simpler in theories in which conformal symmetry is broken spontaneously, so we begin with this case.
The dilaton is then a Goldstone boson of the theory. To find its low-energy effective action, we couple the theory to the general
metric $g_{\m\n}(x)$ and consider the effect of diffeomorphisms and
Weyl transformations
\be
 \lab{weyltrf}
  g_{\m\n} \to g_{\m\n}\, e^{2\s(x)}\,,
  \qquad\qquad
  \t(x) \to \t(x)  + \s(x)\,.
\ee
The dilaton effective action has two classes of terms, those which are
manifestly invariant under \reef{weyltrf} and those which encode the
Weyl anomaly. Of course, both sets of terms are also invariant under
diffeomorphisms.

The anomaly term is more interesting.   It is a
functional $S[g,\t]$ whose linear variation in $\s$ is the trace anomaly
\reef{trace4}. We start with the simple term
 \be \lab{4dsanomtest}
S_\text{tmp}~=~
\int d^4x \sqrt{-g}\ \t\ \left(  c\,
W^2 - a\, E_4 \right)
\,,
\ee
with $W^2$ and $E_4$ as in \reef{trace4}.
The $\d\t(x)$ variation reproduces  the correct Weyl term of \reef{trace4} because the combination $\sqrt{-g}\,W^2$ is Weyl invariant and independent of $\t$. By contrast, $\sqrt{-g}\,E_4$ is not Weyl invariant, so the
variation of the Euler term above yields a spurious term linear in
$\pa_\m\tau$, as well as the desired contribution proportional to
$E_4$. However, the action above can be corrected by adding additional
terms nonlinear in $\tau$ \cite{spon,zohar} such that the final form of the anomalous dilaton Wess-Zumino  action becomes
 \be  \label{4dsanom}
S_\text{anom}
= \int d^4x
\sqrt{-g}\,\Big\{c\,\t \,W^2
 -\,a\Big[\t \,E_4 +4
\big(R^{\m\n} - \tfrac{1}{2} g^{\m\n}R\big)\pa_\m\t\,\pa_\n\t
-4(\pa\t)^2\Box\t + 2 (\pa\t)^4\Big]\Big\}\,.
\ee
In the flat space limit, one can see that \reef{4dsanom} generates order $p^4$ terms in dilaton amplitudes.

We must be careful to include in $S[g,\t]$ all other possible terms in the effective action which
generate amplitudes of order $\le p^4$ and are consistent with the symmetries.
These (Weyl $\times$ diffeo)-invariant terms can
easily be found by writing all independent curvature invariants
 in terms of the manifestly Weyl invariant `metric':
$\hat{g}_{\m\n} = g_{\m\n}\, e^{-2\t}$. Up to four derivatives, the
independent invariants in a conformally flat spacetime are
\cite{zohar}
 \be \lab{4dsinv}
S_\text{inv} = \int d^4x \sqrt{-\hat{g}} \, \Big[ \a - \frac{1}{12}\,f^2\,
\hat{R} +\frac{\k}{36}\, \hat{R}^2 \Big]\,.
\ee
Here $f$ is a constant with mass dimension one.
We now combine eqs.~\reef{4dsinv} and \reef{4dsanom},  take the
flat space limit, \ie $g_{\m\n} \to \h_{\m\n}$, and obtain the total effective
action of the dilaton used in \cite{spon,zohar,Z2}
 \be
 \lab{dilact}
  S_\t
  =
  \int d^4x
  \bigg[ \a\,e^{-4\t} -\frac12 f^2\, e^{-2\t} (\pa\t)^2
    +\k\Big(\Box\t -(\pa\t)^2\Big)^2 + 2a \Big[2 (\pa\t)^2
\Box\t - (\pa\t)^4\Big]\bigg]\,. \ee
Hence Weyl and diffeomorphism invariance  lead to a highly constrained
form for the effective dilaton action even in flat space.

Let us make two further refinements of the action $S_\t$ in order to more easily apply it to the calculation of dilaton scattering amplitudes.
The first step is to `complete the square' in the terms proportional to $a$.
With the shift $\k'=\k-2a$,
we rewrite \reef{dilact}  as
 \be
 \lab{dilact1}
S_\t = \int d^4x \left[\a\,e^{-4\t} -\frac12\,f^2\, e^{-2\t} (\pa\t)^2
+\k'\Big(\Box\t -(\pa\t)^2\Big)^2 +  2a\, \t\,\Box^2\t \right]\,.
 \ee

The second step is the change of
variables
 \be
 \lab{changt}
e^{-\t} = 1 - \frac{\varphi}{f} ,  \quad\qquad\qquad
\t=-\ln\Big(1-\frac{\varphi}{f}\Big) =
\sum_{n=1}^\infty\frac{1}{n}\left(\frac{\varphi}{f}\right)^n\,.
 \ee
The net result is the action
 \be
 \lab{dilact2}
S_{\varphi} = \int d^4x \left[ - \frac12 (\pa\varphi)^2
+\frac{\a}{(1-\varphi/f)^4} +\frac{\k'}{f^2}\frac{(\Box\varphi)^2}{(1-\varphi/f)^2} +2a \sum_{m,n=1}^\infty
\frac{\varphi^m\,\Box^2\varphi^n}{m\,n\,f^{m+n}} \right]\,.
 \ee

Let us now follow the argument of \cite{zohar} to obtain the $a$-theorem for RG flows with spontaneous breaking of conformal symmetry.  The stress tensor in the flat space limit is traceless; $T_\m{}^\m=0$
remains as the Ward identity for global conformal symmetry at the quantum level, although there is an anomaly in curved spacetime.\footnote{A  consequence of the curved space anomaly is that flat space correlation functions involving   $T_\m{}^\m$  contain contact terms.} 
The constant $f$ is related to the VEV of a relevant operator of CFT$_{\rm UV}$ with central charges $a_{\rm UV}$ and $c_{\rm UV}$.  
In the IR, \ie at energies $E \ll f$, where all particles with masses
${\cal O}(f)$ are integrated out, the RG flow ends at a CFT$_{\rm IR}$ with
$a_{\rm IR}$ and $c_{\rm IR}$. The situation is similar for
anomalous chiral symmetries  in which the Ward identity $\pa_\m J^\m =0$ holds in the absence of external sources  whether or not  the symmetries are  spontaneously broken.  If unbroken, the low energy spectrum of the theory contains massless chiral fermions and there is strict anomaly matching,
 $b_\text{UV} = b_\text{IR}$.  If spontaneously broken, the spectrum of the theory contains massless Goldstone bosons, and the strength of their
low-energy self-interactions is fixed by $b_\text{UV}$
 \cite{WZpaper}. For conformal symmetry we have the intermediate situation that the 
total UV and IR anomalies match \cite{spon},  and the difference 
$\D a \equiv a_\text{UV}-a_\text{IR}$ is the coefficient of the dilaton Wess-Zumino term  in the flat-space limit of \reef{4dsanom}. 
 This means that $a$ in \reef{dilact}, \reef{dilact1} and \reef{dilact2} is replaced by
$\D a$.

In principle, all coefficients
of \reef{dilact2}, including $\D a$,
can be calculated by evaluating the path
integral for the interpolating theory whose RG flow is being studied.
The first three coefficients depend on the  renormalization
scheme, but the anomaly coefficient $\D a$
 is universal.  In the present case
of spontaneous breaking of  conformal invariance, there is a moduli space of vacua and thus no potential for the dilaton.  So $\a=0$.

When $\a=0$, it is clear from \reef{dilact2} that $\varphi$ is a canonically normalized field with $\<\varphi\> =0$. It is thus well suited for the computation of  scattering amplitudes and $\varphi$ can be thought of as the `physical dilaton'. The on-shell condition is simply $\Box\varphi=0$ or, equivalently, $p_i^2 =0$ for all external particles of on-shell amplitudes. 
It follows from \reef{dilact2} that
$n$-point
on-shell scattering amplitudes of the
dilaton vanish as $p^4$ at low energy.\footnote{Note that an $n$-point amplitude for any set of identical massless scalars cannot have an order $p^2$ term at low energy because the only  
Bose symmetric Lorentz invariant available vanishes; for example for $n=4$ it is $s+t+u=0$!} For example,  the 4-point amplitude  has low-energy matrix element
 \be \lab{4ptle}
A(s,t) ~=~ \frac{1}{2f^4}\D a\,
\big\<\!-\!p_3,- p_4\big|\,\vf^2\Box^2\vf^2 \, \big|p_1,p_2\big\>
~=~ \frac{4}{f^4}\,\D a\ (s^2 +t^2 +u^2)\,.
 \ee

\begin{figure}
\centerline{\includegraphics[height=6.cm]{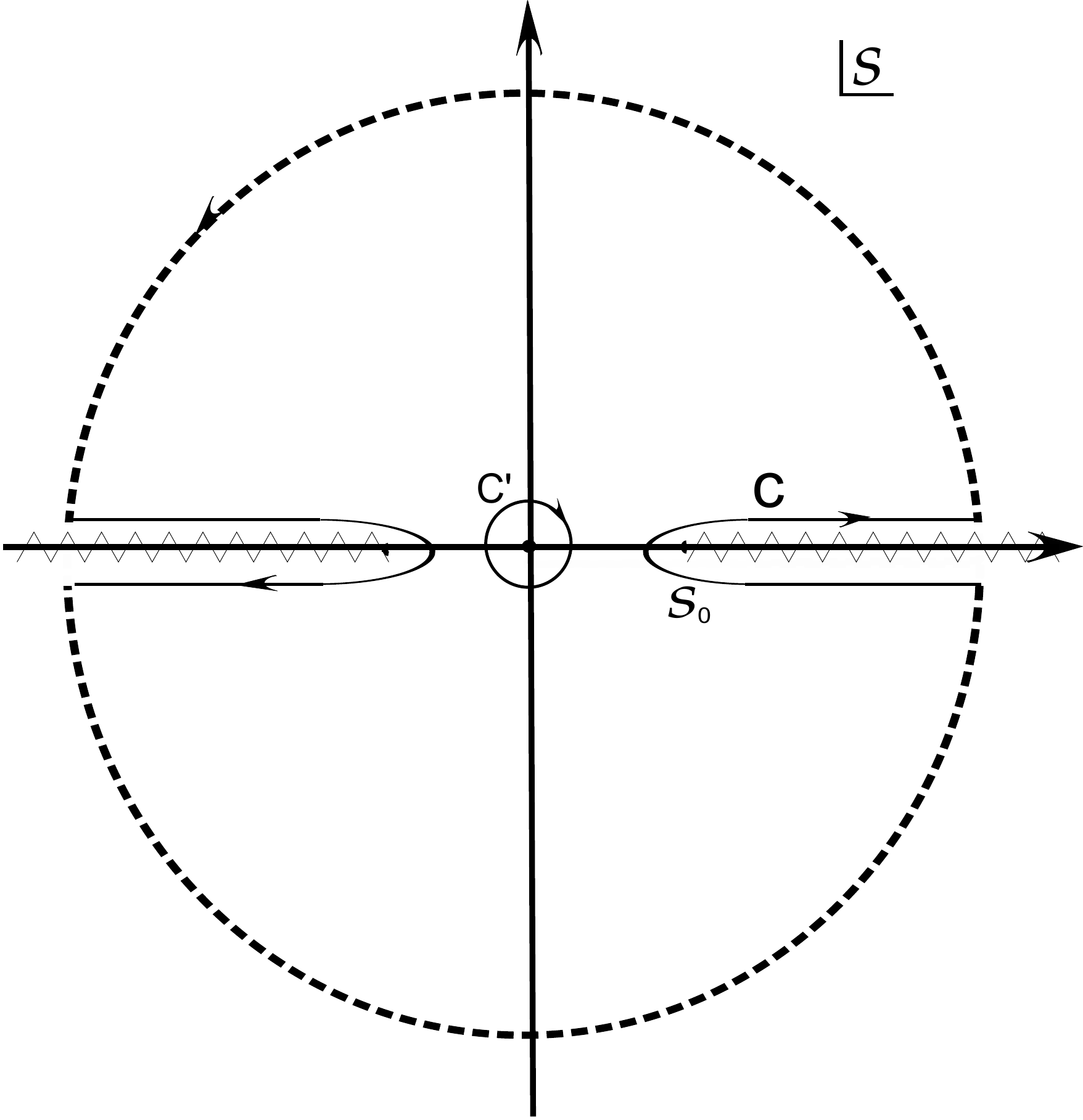}}
\caption{ The contour used to derive \reef{sumrl}  and \reef{contourI}. 
The contour $C'$ surrounds the simple pole of $\ca(s,0)/s^3$ at $s=0$.}
\label{figdisp}
\end{figure}

To prove the $a$-theorem we consider the full amplitude $\ca(s,t)$ which approaches $A(s,t)$ at low-energy. The forward amplitude satisfies a dispersion relation in which the right and left-hand cuts are equal by crossing
symmetry. The simplest way to proceed is to note that
$\ca(s,0)/s^3$ is
analytic in the annular domain which is the interior of the curve
$C\cup C'$ in Fig.~\ref{figdisp}.  Using Cauchy's theorem and crossing,
we obtain
 \be \lab{sumrl}
\frac{8}{f^4}\D a = \frac{2}{\pi}\int_{s_0}^\infty ds \,\frac{{\rm Im}\,\ca(s,0)}{s^3}\,.
 \ee
Here $s_0$ is a temporary IR cutoff which allows us to separate the right and left cuts  from the pole.  In the limit $s_0 \to 0$ we obtain the $a$-theorem
 \be \lab{athm}
a_\text{UV} -a_\text{IR}  = \frac{f^4}{4\pi} \int_{0}^\infty ds\, \frac{{\rm Im}\,\ca(s,0)}{s^3} > 0 \,,
 \ee
where the positive sign is an immediate consequence of unitarity, ${\rm Im}\,\ca(s,0) = s\, \sigma_\text{total}(s)$.
 The interpretation of the sum rule is that the difference $a_\text{UV} -a_\text{IR}$ receives contributions from all scales of the interpolating QFT.

We discuss the sum rule \reef{athm} further below, but we first prefer to bring 
theories with explicitly broken conformal symmetry into the picture. This means that there are explicit scale parameters in the Lagrangian or implicit scales which appear due to dimensional transmutation.  We simply denote these scales by $M$. In this important case, Komargodski and Schwimmer  \cite{zohar}
 introduce the dilaton as a new weakly coupled dynamical field of the theory.
  (Here it can also be viewed as a source \cite{Z2,joep2}.)
 The dilaton kinetic term
of  \reef{dilact} is added with
``decay constant"
$f$ as a free adjustable parameter.
Mass scales are made spacetime dependent by the replacement $M \to M\,e^{-\t(x)}$.  
In other words, the dilaton is added as a conformal compensator. The cosmological or potential term in \reef{dilact2} no longer vanishes.   As in \cite{joep2} we simply add a counter term to cancel it. 
The (improved) stress tensor of the theory modified in this way is traceless, so effectively we have recreated the previous situation of spontaneous breaking  since 
$\Omega(x) =f\, e^{-\t(x)}$ acquires a VEV. With this understanding the previous discussion is applicable, and the sum rule \reef{athm} is derived as above.

There is, however, one important difference between the two cases.  For
RG flows with explicit breaking,  the constant $f$ with dimension of
mass is adjustable. It is chosen to be much larger than any physical
scale \ie  $f \gg M$.  In this limit diagrams containing the dilaton as
an internal line  --- or as an intermediate state in the computation of
${\rm Im}\, \ca(s,0)$ --- are strongly suppressed.
The dilaton then effectively acts as a
source for the trace of the stress tensor in the unmodified theory. For
spontaneously broken flows,  the decay constant $f$ is a fixed physical
constant (it is essentially the VEV),
and effects of the dilaton are included in the low energy
theory.  Indeed, it contributes with the strength of a free massless
scalar   to $a_{\rm IR}$.

It is important to show that the integral in \reef{athm} converges both
at $s=0$ and at large $s$. It is argued in \cite{Z2} and \cite{joep2} that these
limits are controlled by operator deformations of the IR and UV CFT's.
Thus the approach to the
UV is determined by the least relevant operator in the flow away from the
CFT\suv.  Dimensional analysis implies the large $s$ behavior
 \be \lab{uvlim}
 {\rm Im} ~\ca(s,0) \sim s^{2-\e_\text{UV}}\qquad\qquad  {\rm as } ~~s \to \infty\,,
 \ee
with $\e_\text{UV} = {\rm min} (4- \D_i)$.  
The estimate \reef{uvlim} indicates that the contribution of the large circular contour in Fig.~\ref{figdisp} vanishes and that the dispersion integral
converges at high energy.  The  behavior  \reef{uvlim} is confirmed in the massive scalar boson example (in $d=4$) in \cite{Z2}, where   $\e_\text{UV}=2$ and thus $\D=2$ for the mass term $M^2\Phi^2$.  

\begin{figure}
\centerline{\includegraphics[height=4.cm]{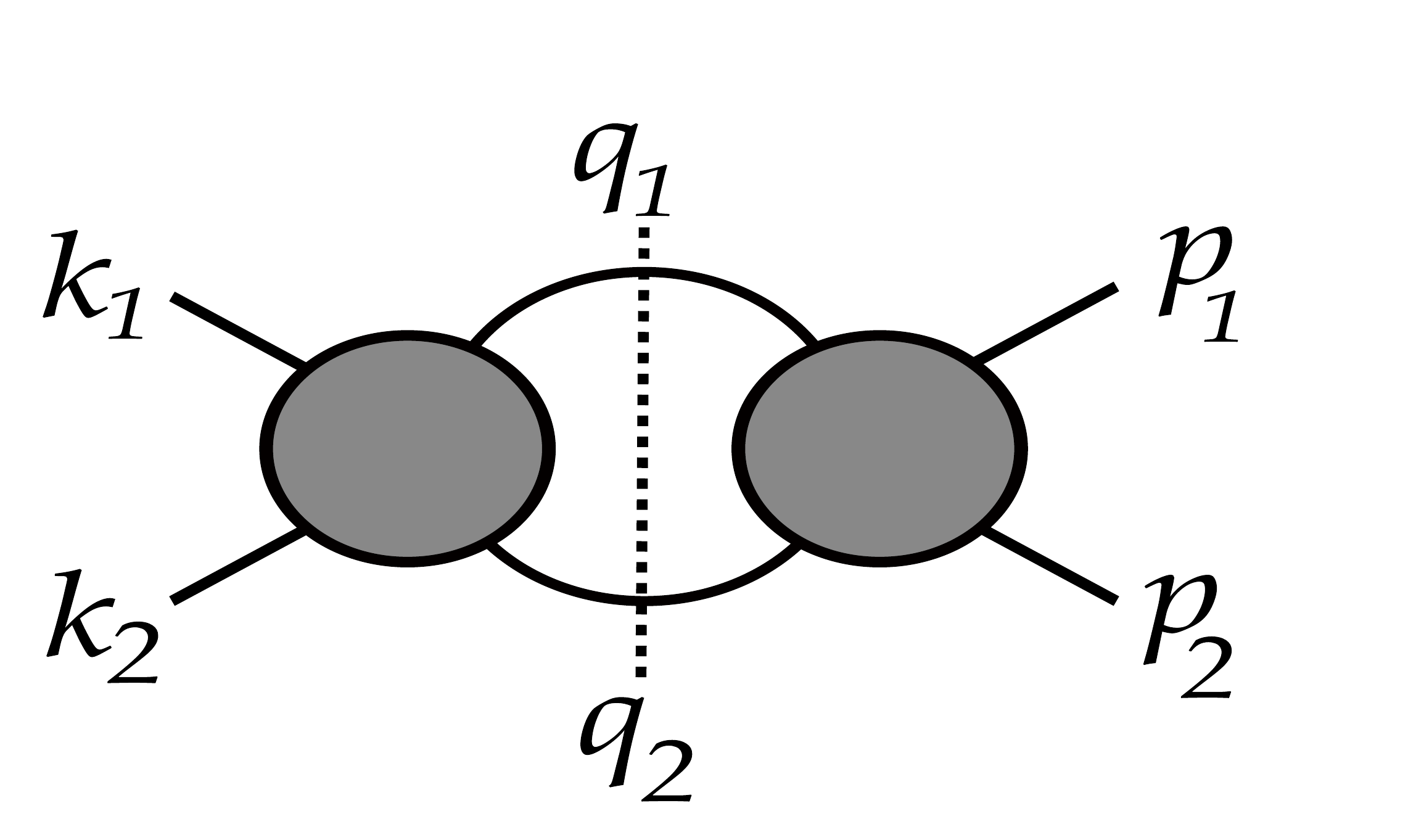}}
\caption{ Computing ${\rm Im} \,\ca(s,t)$ via
unitarity.  The external particles are massless dilatons. In Section 2, the
particles in the intermediate state are also dilatons. In Section 5, they
are massive $\Phi$ particles.
}
\label{figunitarity}
\end{figure}

The approach to the IR should be
determined by the least irrelevant
operator of the CFT\sir, and we expect the low energy behavior
 \be \lab{irlim}
 {\rm Im} ~\ca(s,0) \sim s^{2+\e_\text{IR}}  \qquad \qquad {\rm as }~~ s \to  0\,,
 \ee
with $\e_\text{IR}= {\rm min} ( \D_i-4) >0$.
This makes the integral  \eqref{athm} IR finite.
To exemplify this behavior we discuss an RG flow
with spontaneously broken conformal  symmetry and study the contribution of the 2-dilaton intermediate state to the sum rule \reef{athm} at low energy.  For the Coulomb branch of $\cn=4$ SYM theory, we are interested the 2-dilaton cut of diagrams of the type shown in Fig.~\ref{figunitarity}. Detailed knowledge of the dilaton 4-point amplitude on each side of the cut is not required, all we need to know is that at low energy it behaves as  $\ca(s,t)  \sim (s^2 +t^2+u^2)$  multiplied by a real constant.
Thus in this limit, unitarity tells us that
\be \lab{2dil}
{\rm Im} \ca(s,0)  \sim  \int d\O (s^2 +t'^2 +u'^2)^2  \sim  s^4\,,
\ee
where $t' =-(p_1 -q_1)^2$ and $u'=-(p_1-q_2)^2$.   With the power law $s^4$,
the sum rule \reef{athm} nicely converges at $s =0.$  However, it is not clear to us that it is associated with an irrelevant operator (which would have $\D=6$).

With the extension of these ideas to $d=6$ in view, we distinguish two aspects of the approach of Komargodski and Schwimmer \cite{zohar}:
\begin{itemize}
\item  The Euler action, with coefficient $\D a= a_{\rm UV} - a_{\rm IR}$, determines the    
     form of the dilaton scattering
    amplitudes at low energy in $d=4$. In a given example,  one can, in
    principle,  check this form,  compute the coefficient and compare the result with
    a calculation of $\D a$ by a conventional
    method, such as the heat kernel method \cite{baste}.

\item  They show that  $\D a >0$ in any 4d theory
using the analyticity and unitarity properties of the dilaton amplitude. 
We study whether these properties are as simple and as effective in $d=6$.
\end{itemize}

\setcounter{equation}{0}
\section{Dilaton effective action for $d=6$}
\label{s:dilact}

To discuss the $a$-theorem in $d=6$, our first task  is to derive the dilaton effective action $S_\text{eff}[\tau]$. Following the 4d approach of \cite{spon,zohar}, $S_\text{eff}$ is the most general action whose Weyl-variation $\delta_\s S_\text{eff}$ equals the trace anomaly. In 6d, the trace anomaly $\reef{trace}$ contains  --- in addition to the Euler $a$-anomaly --- three independent Weyl tensor terms $I_i$. We  confine our attention to the Euler anomaly $T_\mu{}^{\mu} = a\, E_6$.  Working out the index contractions in the $2p=6$ case of \reef{term0}, one finds
\bea
E_6  &=&
  \frac{21}{100} R^3 - \frac{27}{20} R\,R_{\m\n}\,R^{\m\n}
 + \frac{3}{2}R^{\m}_{~\,\n}\,R^{\n}_{~\,\l}\,R^{\l}_{~\,\m}
 + \dots
  \, ,
  \labell{6deuler}
\eea
where ``+\dots"
stands for terms that vanish in a conformally flat background; see \reef{e66}.

The Weyl variation $\d_\s E_6$ is non-vanishing, so we need to determine terms $P(\tau,g_{\m\n})$ such that
\bea
  \label{Seuler1}
  S_\text{Euler}
   =  \int d^6 x\,\sqrt{-g}\, \Big(\tau E_6 + P(\tau,g_{\m\n})  \Big)
  ~~~~~\implies ~~~~~
   \delta_\sigma S_\text{Euler}  =  \int d^6 x\,\sqrt{-g} \,\sigma E_6
   \,.
\eea
Then $a \, S_\text{Euler}$ produces the correct anomaly action.
We present two complementary derivations of $S_\text{Euler}$ in
App.~\ref{app:Seuler}. Its flat-space limit is given in \reef{euler6}.

Any Weyl-invariant action can be added to $S_\text{Euler}$ without affecting \reef{Seuler1}. Hence the most general dilaton effective action $S_\text{eff}$ includes also all possible Weyl-invariants.
For our purpose we need to consider all 2-, 4- and 6-derivative Weyl-invariants.\footnote{As in Sec.~\ref{s:D4}, the cosmological term with 0-derivatives is tuned to vanish.}  Since $\hat{g}_{\m\n}=e^{-2\t} g_{\m\n}$ is Weyl invariant, the terms we seek are found by replacing $g_{\m\n} \to \hat{g}_{\m\n}$  in  linear, quadratic, and cubic  curvature invariants.  Since we are interested only in the invariants which remain independent when evaluated on  conformally flat metrics,
\be
 \label{ghat}
\hat{g}_{\m\n}= e^{-2\t}\h_{\m\n}\,,
\ee
 the Riemann tensor $R^\mu{}_{\n\r\s}$ can be replaced by  $W^\mu{}_{\n\r\s}$ plus Ricci-terms (see \reef{weyl}), and we need only consider terms constructed from $\hat{R}_{\m\n}$, $\hat{R}$, and covariant derivatives. Here and henceforth, hatted quantities refer to the conformally flat Weyl-invariant metric \reef{ghat}.
 We classify the possible terms of this type in Secs.~\ref{s:S2deriv}-\ref{s:S6deriv}. The result-oriented reader can skip ahead to \reef{Seff6d} and \reef{Seff6dGEN} which give the form of $S_\text{eff}$ that will be used in the remainder of the paper.

\subsection{2-derivative Weyl-invariants}
\label{s:S2deriv}
In any dimension $d$, there is a unique 2-derivative diffeo-Weyl invariant, namely $\sqrt{-\hat{g}}\,\hat{R}$. It gives rise to the dilaton kinetic term. We present the $d$-dimensional result in \reef{2derb}, but focus here on the $d=6$ case whose flat space limit \reef{ghat} is
\bea
 \label{Rhat}
\sqrt{-\hat{g}}\, \hat{R}
~=~ 10\big(\Box\tau-2(\partial\tau)^2\big) e^{-4\tau}
~\to~20 (\partial\tau)^2\, e^{-4\tau} \,.
\eea
The second expression is obtained by partial integration.
We write the kinetic term of the dilaton as
\bea
  \label{Skin}
  S_\text{kin}
  ~=~  \int d^6x \,
  \bigg[ -  \frac{f^4}{10}\sqrt{-\hat{g}}\, \hat{R} \bigg]
  ~=~ \int d^6x \,\Big[ - 2\, f^4\, (\partial\tau)^2\, e^{-4\tau} \Big]
  \,,
\eea
where $f$ has dimension of mass.
The normalization of the kinetic term is chosen for later convenience.
It follows from \reef{Skin} that the equation of motion of $\tau$ is
\bea
  \label{tauEOM}
  \Box\tau~=~2(\partial\tau)^2 \,.
\eea
It is relevant to note that  $\hat{R} $ is proportional to the $\tau$ equation of motion, as can be seen from \reef{Rhat}.

\subsection{4-derivative Weyl-invariants}
\label{s:S4deriv}

At the level of 4-derivatives, there are three basic curvature invariants: $R^2$, $R_{\m\n} R^{\m\n}$, and $\Box R$. The latter is a total derivative in the effective action, so we do not consider it further. Evaluating the two others  
on the Weyl-invariant metric \reef{ghat} we find
\bea
  \nonumber
  S_{\partial^4}
  &=& \int d^6x \,
  \sqrt{-\hat{g}} \,
  \bigg\{
  b' \, \frac{1}{100} \hat{R}^2
  -b \,\frac{1}{2} \hat{R}_{\mu\n} \hat{R}^{\mu\n}
  \bigg\}
  \\[2mm]
  &=& \int d^6x \,
  \bigg\{
  \,b'\,
    \Big[\Box\tau-2(\partial\tau)^2 \Big]^2 \,e^{-2\tau}
    - b\,
  \Big[
     15 (\Box\tau)^2
     - 68 \Box\tau  (\partial \tau)^2
     + 72  (\partial \tau)^4
  \Big] \,e^{-2\tau}
  \bigg\}
  \,.~~~
    \label{Spa4-no1}
\eea
The couplings $b$ and $b'$ have dimension of (mass)$^{2}$.
It is clear that $\hat{R}^2$ vanishes under the EOM \reef{tauEOM}, but
$\hat{R}_{\mu\n} \hat{R}^{\mu\n}$ does not. It will contribute to scattering amplitudes.

For the purpose of calculating on-shell scattering amplitudes, it is useful to rearrange \reef{Spa4-no1} as
\bea
  S_{\partial^4}
  &=& \int d^6x \,
  \bigg\{
  \,b''\,
    \Big[\Box\tau-2(\partial\tau)^2 \Big]^2 \,e^{-2\tau}
    +4 b\, e^{-\tau} \Box^2 e^{-\tau}
  \bigg\}
  \,~~~
    \label{Spa4-no2}
\eea
with $b'' = b' - 19b$. This follows from using straightforward algebra to show that  $(\Box e^{-\tau})^2$ is a simple linear combination of the two Weyl-invariants in \reef{Spa4-no1}.

\subsection{6-derivative Weyl-invariants}
\label{s:S6deriv}

It is simple to list the general set of 6-derivative Weyl-invariants constructed from Ricci tensors and covariant derivatives. Taking the Bianchi identity and partial integration into account we find
 \bea
  \label{Weylinv6pa}
  \hat{R}^3 \,,~~~~
  \hat{R}\,\hat{R}_{\m\n}\,\hat{R}^{\m\n} \,,~~~~
  \hat{R}\,\hat{\Box}\,\hat{R} \,,~~~~
  \hat{R}^\m_{~\,\n}\,\hat{R}^\nu_{~\,\lambda} \,\hat{R}^\lambda_{~\,\mu} \,,~~~~
  \hat{R}^\m_{~\,\n} \, \hat{\Box}\,\hat{R}^\nu_{~\,\m} \,.
 \eea
This can also be deduced from the list of eleven curvature invariants in
\cite{Bonora} or \cite{HenSken}. Of those, six invariants involve the full  Riemann tensor  which can be replaced by the Weyl tensor plus Ricci's; the 5 remaining invariants are \reef{Weylinv6pa}.

All five invariants vanish when $\tau$ is on-shell; this is clear for the first 3 invariants since they are proportional to $\hat{R}$. As integrated quantities, the latter two are actually not independent from the three others for conformally flat metrics. To see this, first note that the integral of the combination\footnote{
There is a continuation of \reef{ans2} valid for general $d$. See eq.~(3.3) of \cite{anselmi3}.}
\be \lab{ans2}
  20 {R}^\m_{~\,\n} \, {\Box}\,{R}^\nu_{~\,\m}
  -6 {R}\,{\Box}\,{R}
  - 30   {R}^\m_{~\,\n}\,{R}^\nu_{~\,\lambda} \,{R}^\lambda_{~\,\mu} \,
  +11 {R}\,{R}_{\m\n}\,{R}^{\m\n}
  - {R}^3~~~~
\ee
\emph{vanishes for conformally flat metrics}. This is used to eliminate $\hat{R}^\m_{~\,\n} \, {\Box}\,\hat{R}^\nu_{~\,\m}$ in favor of the four other invariants in \reef{Weylinv6pa}. Second, the Euler density $E_6$ is a total derivative, so the integral of $\hat{R}^\m_{~\,\n}\,\hat{R}^\nu_{~\,\lambda} \,\hat{R}^\lambda_{~\,\mu}$ can be expressed in terms of the integral of $\hat{R}^3$ and
$\hat{R}\,\hat{R}_{\m\n}\,\hat{R}^{\m\n}$ via \reef{6deuler}.

Thus the most general Weyl-invariant 6-derivative action can be written as a linear combination of the integrals of $\hat{R}^3$,\, $\hat{R}\,\hat{R}_{\m\n}\,\hat{R}^{\m\n}$, and $\hat{R}\,\hat{\Box}\,\hat{R}$. Since they vanish on-shell, they will not affect the on-shell  dilaton amplitudes at $O(p^6)$, but they are nonetheless useful for us in the following section, so we present their explicit  forms for the conformally flat metric \reef{ghat} in \reef{rcubed}, \reef{rricric}, and \reef{rboxr}.

\subsection{The 6-derivative  Euler action $S_{\rm Euler}$}
\label{s:Seuler}
It is this action whose Weyl variation produces the conformal anomaly.
We have computed $S_{\rm Euler}$ from \reef{Seuler1}
by two related methods, and we refer readers to the self-contained discussions in App.~\ref{app:Seuler}.
For calculations in Secs.~\ref{s:massive}-\ref{s:DBI6d},
 we need only the result in
 the flat space limit,
 \be \lab{euler6}
  S_\text{Euler}
  =
  \int d^6 x\, \Big[
  -24 (\Box\tau)^2 (\pa\tau)^2 + 24 (\pa\tau)^2  (\pa\pa\tau)^2
  + 36 \Box\tau  (\pa\tau)^4
  - 24 (\pa\tau)^6
  \Big] \,.
 \ee
Amplitudes are easier to calculate after a rewriting of the action \reef{euler6}. It takes straightforward algebra to show that
\bea
  S_\text{Euler}
  ~=~
  \int d^6 x\,
  \bigg[
   \sqrt{-\hat{g}}\,
    \bigg(
     \frac{39}{1000} \, \hat{R}^3
     - \frac{3}{20}\, \hat{R}\,\hat{R}_{\m\n}\,\hat{R}^{\m\n}
     - \frac{3}{100}\, \hat{R}\,\hat{\Box}\,\hat{R}
     \bigg)
     ~+~ 3 \,\tau\Box^3 \tau
  \bigg] \,,  \lab{mapleleaf}
\eea
where all hatted terms are evaluated on the conformally flat metric \reef{ghat}. 
The first three terms can be absorbed in the general 6-derivative Weyl-invariant terms discussed in Sec.~\ref{s:S6deriv}, and we can then write the most general 6-derivative effective action as
 \be \lab{euler6-new}
  S_{\pa^6}
  =
  \int d^6 x\,
  \bigg[
   \sqrt{-\hat{g}}\,
    \Big(
     \,b_1 \, \hat{R}^3
     + b_2\, \hat{R}\,\hat{R}_{\m\n}\,\hat{R}^{\m\n}
     + b_3\, \hat{R}\,\hat{\Box}\,\hat{R}
     \Big)
     ~+
       ~3 a \,\tau\Box^3 \tau\,
  \bigg] \,.
 \ee
Note that  \reef{mapleleaf} requires an algebraic miracle:  7 equations
in 3 unknowns have a unique solution.  We refer  readers who believe in
mathematics rather than miracles to the end of App.~\ref{app:Seuler}, 
where the formalism of Paneitz operators and $Q$-curvatures is discussed briefly.

When $\tau$ satisfies the equation of motion $\Box\tau = 2 (\pa \t)^2$, the three Weyl-invariants in \reef{euler6-new} vanish. Thus only the last term, $3 a \,\tau\Box^3 \tau$, contributes to the scattering amplitudes and it captures the information about the $a$-anomaly. Repeated use of the equation of motion (and partial integration) shows that  $3\tau \Box^3 \tau \to
\Big[ 24 (\pa\tau)^2  (\pa\pa\tau)^2 - 48 (\pa\tau)^6 \Big]$. This latter is  the same result obtained by  applying the equations of motion in \reef{euler6}.

\subsection{The dilaton effective action}
\label{s:Seffsummary}
The 4d argument of Komargodski and Schwimmer \cite{zohar}, as reviewed in
Sec.~\ref{s:D4}, shows that the change in the Euler central charge in the flow  from the CFT$_\text{UV}$  to  CFT$_\text{IR}$  will be carried by the dilaton.
Our discussion in this section can  then be summarized in the flat-space limit of the dilaton effective action 
\be
   \boxed{
   \begin{array}{rcl}
  S_\text{eff}
  &=&
  \displaystyle \int d^6 x \,\bigg[ - 2 f^4 (\partial\tau)^2\, e^{-4\tau}
   +  b''\,
    \Big[\Box\tau-2(\partial\tau)^2 \Big]^2 \,e^{-2\tau}
  + 4 b\, e^{-\tau} \Box^2 e^{-\tau} \\[2mm]
  &&
  \hspace{1.5cm}
    + \sqrt{-\hat{g}}\,
    \Big(
     \,b_1 \, \hat{R}^3
     + b_2\, \hat{R}\,\hat{R}_{\m\n}\,\hat{R}^{\m\n}
     + b_3\, \hat{R}\,\hat{\Box}\,\hat{R}
     \Big)
  ~+~
  3 \Delta{a} \,\tau\Box^3 \tau \bigg] \,,
  \end{array}}
    \label{Seff6dGEN}
\ee
with expressions for the 6-derivative Weyl-invariants given in \reef{rcubed}-\reef{rboxr}. Dropping terms that vanish on the $\tau$ equations of motion and therefore do not contribute to the low-energy on-shell dilaton amplitudes of our interest, we can simplify this to
\bea
  S_\text{eff}
  &=&
  \int d^6 x \,\bigg[ - 2 f^4 (\partial\tau)^2\, e^{-4\tau}
  + 4 b\, e^{-\tau} \Box^2 e^{-\tau}
  +
  3 \Delta{a} \,\tau\Box^3 \tau \bigg] \,.
  \labell{Seff6d}
\eea
Compared with 4d, it is a  new feature in $d=6$ that  both a Weyl$\times$diffeo invariant 4-derivative term and the 6-derivative Euler anomaly term contribute to on-shell dilaton amplitudes.

\subsection{Dilaton matrix elements from the effective action}
\label{s:varphi}
We are interested in the on-shell matrix elements of \reef{Seff6d} 
at 4th and 6th order in the
low-energy
small-momentum expansion. As discussed in Sec.~\ref{s:D4},  we  first  transform  to  the `physical dilaton' field
$\varphi$,  defined in 6d by
\be
\label{tau-varphi-6d}
e^{-2\t} ~=~ 1 - \frac{\varphi}{f^2} ,
\quad\qquad\qquad
\t~=~-\frac{1}{2}\ln\Big(1-\frac{\varphi}{f^2}\Big)
~=~
\frac{\varphi}{2f^2}
+\frac{\varphi^2}{4f^4}
+ \dots
\,.
\ee
The $\t$ equation of motion \reef{tauEOM} implies $\Box\varphi = 0$, and the on-shell condition for physical dilaton is therefore simply $p^2=0$.

From the action \reef{Seff6d} we obtain\footnote{Note that we have dropped 6-derivative terms which do not contribute to the $O(p^6)$ on-shell amplitudes, for example $\varphi^3 \Box^3 \varphi$.  We have also dropped quadratic terms $\varphi \Box^2 \varphi$ and $\varphi \Box^3 \varphi$ which do not influence the amplitudes of interest here.}
\bea
  \nonumber
  S_\text{eff}  &=& \int d^6x \,
  \bigg\{
  -\frac{1}{2}(\pa\varphi)^2
\\[2mm] \nonumber
&& \hspace{1.6cm}
  + \frac{b}{2 f^6} \,\varphi^2 \Box^2 \varphi
  + \frac{b}{16 f^8}
     \Big[ \,4\varphi^3 \Box^2 \varphi + \varphi^2 \Box^2 \varphi^2 \Big]
  + \frac{b}{32 f^{10}}
     \Big[ \,5 \varphi^4 \Box^2 \varphi + 2 \varphi^3 \Box^2 \varphi^2 \Big]
     \\[2mm]
    &&\hspace{1.6cm}
     + \frac{b}{128 f^{12}}
     \Big[ \,14\,\varphi^5 \Box^2 \varphi + 5 \,\varphi^4 \Box^2 \varphi^2
     + 2\,\varphi^3 \Box^2 \varphi^3
     \Big]
  \labell{S-phi}
     \\[2mm] \nonumber
     &&\hspace{1.6cm}
     +
       ~\frac{3\Delta{a}}{16f^8} \,\varphi^2 \Box^3 \varphi^2
       +\frac{\Delta{a}}{4f^{10}} \,\varphi^3 \Box^3 \varphi^2
       +\frac{\Delta{a}}{48 f^{12}} \,
       \Big[ 9 \varphi^4 \Box^3 \varphi^2
       + 4 \varphi^3 \Box^3 \varphi^3
       \Big]
     + O(\varphi^7)
  \bigg\}
  \,.~~
\eea
Next we extract the dilaton matrix elements.

\vspace{2mm}
\noindent {\bf Matrix elements at $O(p^4)$}\\[1mm]
It is easy to extract the on-shell matrix elements from \reef{S-phi}. At order $O(p^4)$, the results can be read off directly from the interaction vertices. Any vertex with $\Box\varphi$ can be dropped.  The term  $(\Box)^k \vf^m$ with $k >2$
contributes simply as  $m! (s_{i_1 i_2 \ldots i_m})^k$,  where 
 $s_{i_1 i_2 \ldots i_m} = -(p_{i_1}+p_{i_2}+\ldots +p_{i_m})^2$.
For example, for the 4-point matrix elements, the only contribution comes from
$\varphi^2 \Box^2 \varphi^2$ which gives the vertex rule $8(s^2 + t^2 +u^2)$.
We list the results at $O(p^4)$ below for later reference:
\bea
  \nonumber
  A_4^{O(p^4)} &=& \frac{b}{2 f^8}(s^2 + t^2 +u^2) \, ,\\[2mm]
  \label{Amp-p4th}
  A_5^{O(p^4)} &=& \frac{3b}{4 f^{10}}  \sum_{1\le i<j\le 5} s_{ij}^2\, , \\[2mm]
  \nonumber
  A_6^{O(p^4)} &=& \frac{3b}{f^{12}}  \sum_{1\le i<j\le 6} s_{ij}^2\, .
\eea
In the last case, we used that $\sum_{1\le i<j<k\le 6} s_{ijk}^2 =2 \sum_{1\le i<j\le 6} s_{ij}^2$ for null momenta.

\vspace{2mm}
At  $O(p^6)$ the matrix elements receive contributions from the 6-derivative contact terms in \reef{S-phi}, but if the dilaton is dynamical --- as in the spontaneously broken case --- there will also be contributions from pole diagrams with two vertices from 4-derivative terms in \reef{S-phi}. We will consider these two cases separately.

\compactsubsection{Matrix elements at $O(p^6)$: {\em explicitly broken conformal symmetry}}
When conformal symmetry is softly broken by relevant operators, we can regard the dilaton as a weakly coupled scalar with the scale $f$ chosen much larger than any other scale in the problem; in particular $f^2 \gg b$, so since pole diagrams at order $p^6$  are $O(b^2/f^4)$ they will be suppressed. Alternatively, we can regard the dilaton as a source and in that case there are simply no pole diagrams.
The on-shell $O(p^6)$ matrix elements are found directly from the $O(\pa^6)$ contact terms in \reef{S-phi} and they are
\bea
  \label{ME4ptp6}
  A_4^{O(p^6)} &=&
  \frac{3\Delta{a}}{2f^8}(s^3 + t^3 +u^3)
  ~=~
  \frac{9\Delta{a}}{2f^8}
    \,s\, t\, u  \, ,
\\[2mm]
    \label{ME5ptp6}
  A_5^{O(p^6)} &=&
      \frac{3\Delta{a}}{f^{10}} \sum_{1\le i<j\le 5} s_{ij}^3\,,\\[2mm]
  A_6^{O(p^6)} &=&
      \frac{3\Delta{a}}{f^{12}}
  \bigg[
  ~3 \sum_{1\le i<j\le 6} s_{ij}^3
  ~~~+~ \sum_{1\le i<j<k\le 6} s_{ijk}^3\,
  \bigg]\,.
    \label{ME6ptp6}
\eea
The simplest example of this case is the free massive scalar which we study in   Sec.~\ref{s:massive}.

\compactsubsection{Matrix elements at $O(p^6)$: {\em spontaneously broken conformal symmetry}}
When the conformal symmetry is broken spontaneously,
 the dilaton is the corresponding Goldstone boson and as such it is a dynamical degree of freedom of the theory. Therefore we must include the pole diagrams with dilaton exchanges. For example, the 4-point matrix element receives a contribution from the tree-diagram with two $O(p^4)$ 3-vertices from \reef{S-phi}. The value of the $s$-channel diagram is
\bea \label{tree-diag}
  \raisebox{-5mm}{\includegraphics[height=1.3cm]{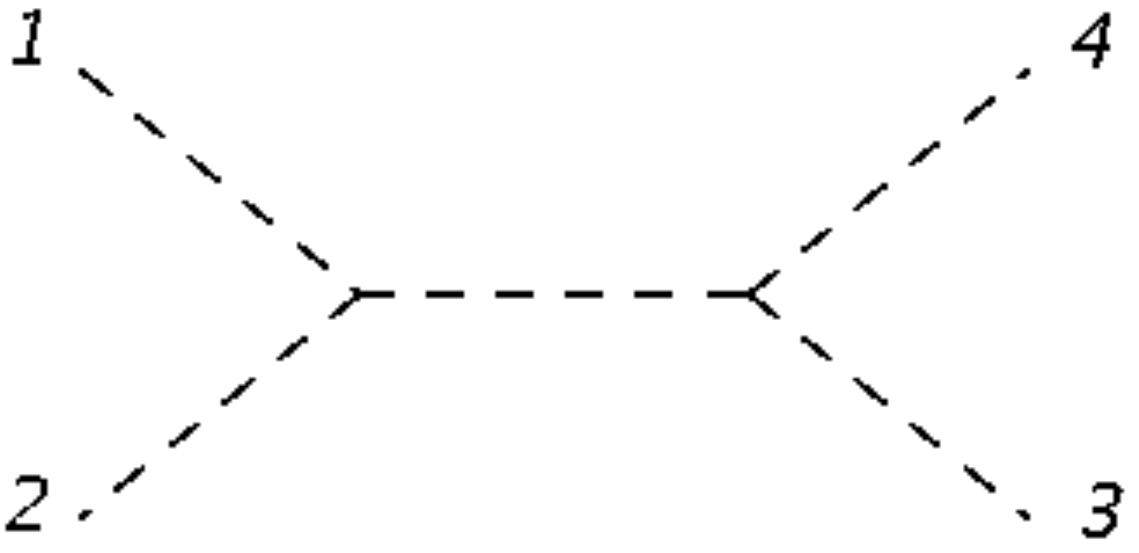}}
  ~=~ \frac{b}{f^6} s^2\times \frac{-1\,}{\,\,s} \times s^2 \frac{b}{f^6}
  ~=~-\frac{b^2}{f^{12}}\, s^3
  \, .
\eea
Adding to this the $t$-channel and $u$-channel diagrams, it is clear that the contribution from these diagrams are local and of exactly the same form, $s^3 + t^3 + u^3$, as the contact term contribution that gave \reef{ME4ptp6}. Similarly, the 5- and 6-point matrix elements receive contributions from 2- and 3-particle channel tree diagrams. We simply list the results
\bea
  \label{AO4p6}
  A_4^{O(p^6)} &=&
  \bigg[
    \frac{3}{2}\Delta{a} - \frac{b^2}{f^4}
  \bigg]
  \frac{3 }{f^8}\,s\, t\, u  \, ,
  \\[3mm]
  \label{AO5p6}
   A_5^{O(p^6)} &=&
   \bigg[
     \frac{3}{2}\Delta{a}
      - \frac{b^2}{f^{4}}
   \bigg]
   \frac{2}{f^{10}}
      \sum_{1\le i<j\le 5} s_{ij}^3\,,
      \\[3mm]
  \nonumber
        A_6^{O(p^6)}
  &=&
    \bigg[
     \frac{3}{2}\Delta{a}
      - \frac{b^2}{f^{4}}
   \bigg]
   \frac{6}{f^{12}}
      \sum_{1\le i<j\le 6} s_{ij}^3
   +
   \bigg[
    \Delta{a}
      - \frac{5b^2}{8f^{4}}
   \bigg]
   \frac{3}{f^{12}}
      \sum_{1\le i<j<k\le 6} s_{ijk}^3   \\[3mm]
    &&
   ~-~
   \frac{b^2}{4f^{4}}
   \frac{1}{f^{12}}
   \bigg(
       \big(s_{12}^2+s_{13}^2+s_{23}^2\big)
   \times \frac{1}{s_{123}}\times
       \big(s_{45}^2+s_{46}^2+s_{56}^2\big)  + \text{perms}
   \bigg)
      \,.~
    \label{AO6p6}
\eea
Obviously, for $b \ll f^2$, the matrix elements in \reef{AO4p6}-\reef{AO6p6} reduce to those of \reef{ME4ptp6}-\reef{ME6ptp6}.
The 6-point matrix elements also have terms with poles in $s_{ijk}$;  
the ``perms" indicate  the sum of 10 independent 3-particle channels.
It is worth noting that the combination
$\Big[ \frac{3}{2}\Delta{a} - \frac{b^2}{f^{4}}\Big]$
shows up in all three $O(p^6)$ matrix elements.
This plays a role in our study of the 6d (2,0) theory in Sec.~\ref{s:DBI6d}.

\setcounter{equation}{0}
\section{Example of explicit breaking: a free massive scalar}
\label{s:massive}

As the simplest example of explicitly broken conformal symmetry, we take the CFT$_\text{UV}$ to be the 6d theory of a free massless scalar $\Phi$ and deform it with a mass term operator $-\tfrac{1}{2} m^2 \Phi^2$. In the far IR, the massive field $\Phi$ decouples and the CFT$_\text{IR}$ is trivial with no degrees of freedom. Hence $a_\text{IR} =0$. The $a$-central charge in the UV is that of a conformally coupled 6d free massless scalar,
 \begin{equation}\label{asc}
    a_\text{UV} ~=~  a_{\rm sc}~=~\frac{1}{(4 \pi)^3 2^4 3^4 7}
     ~=~
     \frac{1}{(4 \pi)^3\, 9072} \,.
\end{equation}
This value was computed in \cite{baste} using heat kernel methods.
For this RG flow, we therefore have
$\Delta a =  a_\text{UV} - a_\text{IR} = a_{\rm sc}$.

Following \cite{zohar},  we introduce the dilaton via the conformal compensator
$\Omega=f^2 e^{-2\tau}$, where $f$ is a mass scale that we choose such that
$f \gg m$. Specifically, we start with
\bea
  S=-\int d^6x \,\Big( \frac{1}{2} (\pa \Phi)^2 +  \frac{1}{2} m^2 \Phi^2  \Big)\,
\eea
and introduce $\O$ as a canonically normalized scalar of mass dimension 2 such that the coupled action is
\begin{equation}
      S=-\int d^6x \,\Big( \frac{1}{2} (\pa \Phi)^2 +\frac{1}{2} (\pa \Omega)^2 +\frac12 \l \,\Omega\, \Phi^2\Big) \,.
   \label{action1}\,
\end{equation}
The model is weakly coupled since the dimensionless coupling $\l \equiv m^2/f^2$ is small thanks to $f \gg m$, and the  action \reef{action1} has a traceless improved stress tensor. For $\<\Phi\>=0$, there is a moduli space in $\Omega$, and the model is conformally invariant at the origin of moduli space where $\< \Omega\> =0$. When $\Omega$ acquires a non-zero VEV, $\< \Omega\>= f^2$, the conformal symmetry is spontaneously broken, and the fluctuation $\varphi$ around the VEV,
$\Omega =f^2 - \varphi$, is the associated Goldstone boson. In fact, we recognize $\varphi$ as the physical dilaton introduced in \reef{tau-varphi-6d}.
The action coupling $\varphi$ to $\Phi$ is
\bea
  S=-\int d^6x \,\Big( \frac{1}{2} (\pa \Phi)^2 +\frac{1}{2} (\pa \varphi)^2
  + \frac{1}{2} m^2\, \Phi^2
  - \frac{1}{2} \lambda\,\varphi\, \Phi^2
  \Big) \, .
  \label{SwtihVEV}
\eea

Upon integrating out the massive field $\Phi$, the action \reef{SwtihVEV} should yield the dilaton effective action. To verify this, it suffices to show that the on-shell $\varphi$ amplitudes calculated at low-energy from \reef{action1} agree with those of the dilaton effective action, as computed in Sec.~\ref{s:varphi}.
Thus, taking advantage of $\l = m^2/f^2 \ll 1$, we proceed perturbatively and  calculate the 4-, 5-, and 6-point amplitudes of the dilaton $\varphi$ from 1-loop diagrams with $\Phi$ fields on internal lines, extracting their order $p^4$ and $p^6$  terms in the low-energy expansion.

In the action \reef{SwtihVEV}, $\varphi$ is coupled to $\Phi$ only through the cubic interaction term $- \tfrac{1}{2} \lambda\,\varphi\, \Phi^2$. Therefore it is quite simple to calculate the 1-loop dilaton scattering amplitudes. For example, to obtain the on-shell 4-point function of $\varphi$, we have to
 sum 3 permutations of the elementary box diagram
\bea
  I^{1234}_4 = \lambda^4 \int \frac{d^6 \ell}{(2\pi)^6} \frac{1}{\big(\ell^2+m^2\big)\big((\ell+p_1)^2 + m^2 \big)\big((\ell+p_1+p_2)^2 + m^2 \big)\big((l-p_4)^2 +m^2\big)} \, .
\eea
Computing this diagram and its permutations using Feynman parameters, we obtain the following low-energy expansion $s,t,u \ll m^2$ for the 4-point amplitude:
\begin{equation}\lab{A4p4}
    \mathcal{A}_4(\varphi_1\varphi_2\varphi_3\varphi_4)
    ~=~I^{1234}_4+I^{1243}_4+I^{1423}_4
    ~=~\frac{1}{(4\pi)^3f^8}\Biggl[\frac{s^2+t^2+u^2}{720}m^2+\frac{stu}{2016}\Biggr]+O(p^8)\,.
\end{equation}

The calculation of  5-point functions requires the sum of 12 independent pentagon diagrams.  The low energy contribution can be expressed in terms of the 5-point invariants
\bea
\label{A5p4}
~~A_5^{O(p^4)}~=~\frac{m^2}{f^{10}} \frac{1}{(4\pi)^3}
   \, \frac{1}{480} \,\sum_{1\le i<j\le 5} s_{ij}^2 \,,~~\\[2mm]
\label{A5p6}
 ~~A_5^{O(p^6)}~=~\frac{1}{f^{10}} \frac{1}{(4\pi)^3}
   \, \frac{1}{3024} \,\sum_{1\le i<j\le 5} s_{ij}^3 \,.~~
\eea

Proceeding analogously to 6-point functions, we find that the sum of 60 independent hexagon diagrams has the low energy expansion
\bea
\label{A6p4}
 ~~A_6^{O(p^4)}&=&\frac{m^2}{f^{12}} \frac{1}{(4\pi)^3}
   \, \frac{1}{120} \,\sum_{1\le i<j \le 6} s_{ij}^2 \,,~~\\[2mm]
\label{A6p6}
  ~~A_6^{O(p^6)}&=&\frac{1}{f^{12}} \frac{1}{(4\pi)^3}
   \, \frac{1}{3024} \,  \bigg[
  ~3 \sum_{1\le i<j\le 6} s_{ij}^3
  ~~~+~ \sum_{1\le i<j<k\le 6} s_{ijk}^3\,
  \bigg]
  \,.~~
\eea

Let us now compare these amplitudes with matrix elements of Sec.~\ref{s:varphi}.
The $O(p^4)$ results \reef{A4p4}, \reef{A5p4}, and \reef{A6p4} completely match the matrix elements in  \reef{Amp-p4th} with a unique and consistent identification of $b$ as
\bea
 \label{bmsq}
 b=\frac{m^2}{360 (4\pi)^3}\,.
\eea
Next, at  $O(p^6)$, the amplitudes \reef{A4p4}, \reef{A5p6}, and \reef{A6p6} are fully consistent with the dilaton matrix elements \reef{ME4ptp6}, \reef{ME5ptp6}, and \reef{ME6ptp6} with the identification of $\Delta{a} = 1/((4 \pi)^3\, 9072) =a_\text{sc}$. This is in agreement with the expectation, as discussed below \reef{asc}.

We conclude that the simple example of a free massive scalar confirms the structure of the dilaton effective action derived in Sec.~\ref{s:dilact}. Note that it is a key point in the analysis that $f$ can be chosen freely, in particular such that $f \gg m$. This ensures that the amplitudes can be calculated perturbatively in small $\lambda = m^2/f^2$. Moreover, $b^2/f^4 \sim \lambda^2 \ll 1$ shows that the pole exchange diagram contributions in \reef{AO4p6}-\reef{AO6p6} are 2-loop effects in this example, and they are arbitrarily suppressed.

\setcounter{equation}{0}
\section{The anomaly from dispersion relations}
\label{s:dispersiona} 

One of the most striking features of the approach to the $a$-theorem in \cite{zohar} is the use of a dispersion relation  for the forward 4-point dilaton scattering amplitude. Unitarity then provides a quick  proof that  $a_\text{UV}  - a_\text{IR} > 0$ for a general RG flow in $d=4$.  It is an obvious question to ask if a similar approach can work in $d=6$. We begin with 
the 4-point amplitude for an RG flow with explicit breaking, then discuss spontaneous breaking, and finally briefly comment on 6-particle dispersion relations. 

\compactsubsection{Explicit breaking of conformal symmetry}
The low-energy expansion of the 4-dilaton amplitude is given by \reef{Amp-p4th} and \reef{ME4ptp6}:
\begin{equation}\lab{loweng}
  \ca_4(s,t) ~=~\frac{b}{2f^8}(s^2+t^2+u^2)
  +\frac{9}{2f^8} \Delta a\, stu
~+~O(p^8)\,.
\end{equation}
with $s+t+u=0$ since the external dilatons are massless.
Since $\ca_4(s,0) = \tfrac{b}{f^8} s^2+O(s^4)$, a proof that  $\Delta{a}>0$ based on the forward limit  $t\to0$ of the 4-point amplitude cannot be constructed as was done in 4d. 

Instead we derive a dispersion relation for the forward limit of the $t$-derivative of the 4-dilaton amplitude, noting that 
$\tfrac{\pa}{\pa t} \ca''(0,0) =  - \tfrac{9}{f^8}\,\Delta{a}$, where the primes denote $s$-derivatives taken at constant $t$. 
We begin with the contour integral
\be
 \label{contourI}
  I  = \int_{C\cup C'} \frac{ds}{2\pi i}\frac{\mathcal{A}_4(s,t)}{s^3} =0 \,,
\ee
where the contours are sketched in Fig.~\ref{figdisp}.
We take $t$ to be small and negative. The integral vanishes by Cauchy's theorem because there are no singularities enclosed
by the contour.
The integral over the circle $C'$ at $s=0$ gives $\frac{1}{2}\mathcal{A}''(s\!=\!0,t)$.  The horizontal contour contains the contributions from the $s$- and $u$-channel branch cuts. The $s-$ and  $u-$channel branch points are at $4m^2$ and $-t-4m^2$, respectively.  The latter follows from $s+t+u=0$. 
(In the example of the free massive scalar, $m$ is the mass  of the field  $\Phi$.) Equation \reef{contourI} then tells us that
\begin{equation}\label{step1}
      \frac{1}{2} \ca_4''(0,t)~=~
  \int_{4m^2}^\infty\frac{d s }{\pi\,s^3} \,\text{Im}\,\mathcal{A}_4(s,t)
  + \int_{-\infty}^{\dash t\dash 4m^2}\!\!\!\frac{d s }{\pi\,s^3} \,\text{Im}\,\mathcal{A}_4(s,t)\,.
\end{equation}
Crossing symmetry  (specifically $s \lra u$ at fixed $t$)  gives
${\cal A}_4(s+i\epsilon,t)={\cal A}_4(\!-\!s\!-\!t\!-\!i\epsilon,t)$,  so we rewrite the second integral,
\begin{equation}
    \int_{-\infty}^{\dash t\dash 4m^2}\!\!\!\frac{d s }{\pi\,s^3} \,\text{Im}\,\mathcal{A}_4(s,t)
    ~=~ - \int_{4m^2}^\infty\frac{d u }{\pi\,(-t-u)^3} \,\text{Im}\,\mathcal{A}_4(u,t)\,.
\end{equation}
After a trivial relabeling of integration variable $u \to s$, we can combine it 
with the first integral in~(\ref{step1}) to find
\begin{equation}\label{disp1}
      \frac{1}{2}\ca_4''(0,t)~=~
  \int_{4m^2}^\infty\frac{d s }{\pi} \biggl[\frac{1}{s^3}+\frac{1}{(t+s)^3}\biggr]\,\text{Im}\,\mathcal{A}_4(s,t)\,.
\end{equation}
The anomaly coefficient $\Delta a$  is proportional to the $t$-derivative of $A_4''(s\!=\!0,t)$:
\begin{equation}
    \Delta a~=~- \frac{f^8}{9}\frac{\partial}{\partial t}A_4''(s,t)\Bigr|_{s,t=0}\,.
\end{equation}
We now use~(\ref{disp1}) to express $\Delta a$ as the dispersion integral
\begin{equation}\label{deltaadisp}
      \Delta a~=~
  \frac{2f^8}{9\pi}\int_{4m^2}^\infty \,d s\biggl[\frac{3}{ s^4}\text{Im}\,\mathcal{A}_4(s,0)-\frac{2}{s^3}\frac{\partial}{\partial t}\text{Im}\,\mathcal{A}_4(s,t)|_{t=0}\biggr]\,.
\end{equation}

We can learn more about the two terms in the integrand from their 
partial wave expansions.  In $d$-dimensions they are \cite{giddings}
\bea
\ca_4(s,t)  &=& \psi_\l\,s^{2 - \frac{d}{2}} \sum_{n=0}^\infty (n+\l) C^\l_n(\cos(\theta)) a_n(s)\,,\\
\frac{\pa \ca(s,t)}{\pa t}\Big|_{t= 0}&=&  2   \psi_\l\,s^{1 - \frac{d}{2}}   \sum_{n=0}^\infty \frac{n(n+2\l)}{2\l+1}(n+\l) C^\l_n(1) a_n(s) \,,
\eea
with $\l = (d-3)/2$ and $\psi_\l =2^{4\l+3}\pi^\l\G(\l)$.  In the second line we used a property of the derivative of the Gegenbauer polynomials $C^\l_n(z)$. Partial wave unitarity, \ie 
Im\,$a_n(s) \ge |a_n(s)|^2$,  implies that  Im $\ca_4(s,0)$ and  Im $ \pa_t\ca_4(s,t)|_{t=0}$ are both \emph{positive}. Unfortunately, they enter the dispersion relation \reef{deltaadisp} with \emph{opposite signs}!

\begin{figure}
\centerline{
 \includegraphics[height=4.4cm]{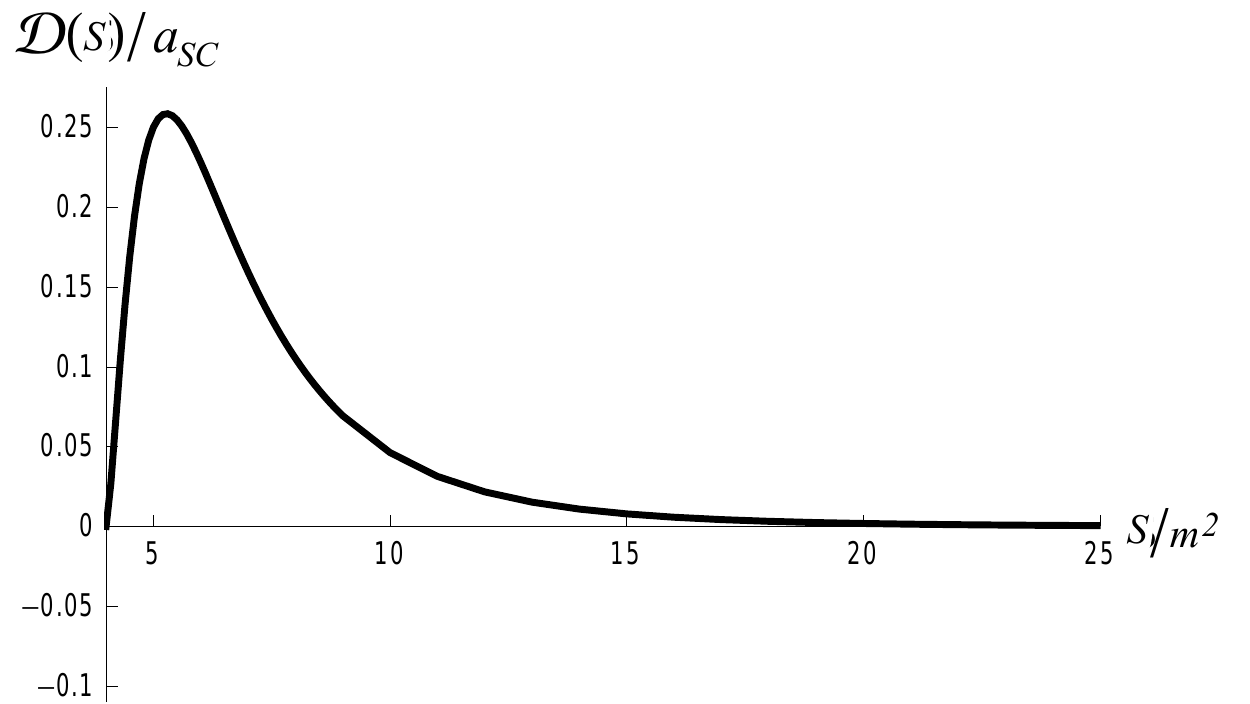}
\hspace{5mm} \includegraphics[height=4.4cm]{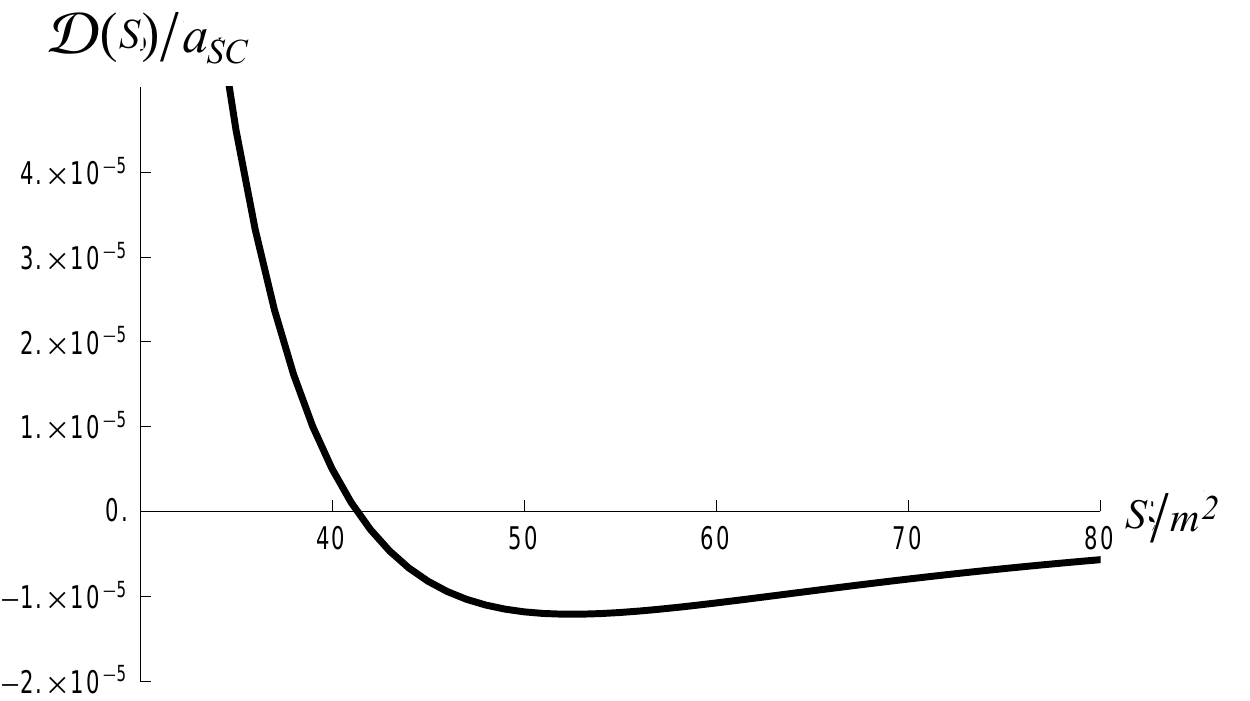}~
}
\caption{The integrand ${\cal D}(s)$ of the dispersion relation relation normalized in units of the scalar anomaly $a_{\rm sc}$.}
\label{figDs}
\end{figure}

Thus it is certainly not  obvious from \reef{deltaadisp} whether $\Delta a$ is positive. It therefore becomes instructive to  compute the integrand of this sum rule in the free massive scalar example.
We use  unitarity to compute $\text{Im}\,\mathcal{A}_4(s,t)$. Let us denote the incoming and outgoing momenta in $\mathcal{A}_4(s,t)$ by $k_1,k_2$ and $p_1,p_2$, respectively.\footnote{Although elsewhere in this paper we take all momenta outgoing, we  work here with in- and outgoing momenta.} Furthermore, we denote the momenta of the two internal on-shell scalars by $q_1,q_2$. See Figure~\ref{figunitarity}.
We then define Mandelstam invariants
\begin{equation}
\begin{split}
    s &= -(k_1+k_2)^2\,,\qquad  t=-(k_1-p_1)^2\,,\qquad u=-(k_1-p_2)^2\,,\\
    s'=s &= -(q_1+q_2)^2\,,\qquad  t'=-(k_1-q_1)^2\,,\qquad u'=-(k_1-q_2)^2\,,\\
    s''=s &= -(p_1+p_2)^2\,,\qquad  t''=-(p_1-q_1)^2\,,\qquad u''=-(p_1-q_2)^2\,.
\end{split}
\end{equation}
The internal $\Phi$-lines are massive of mass $m$,  
while the external dilaton $\varphi$-lines are massless. 
Therefore, $s+t+u=0$ while $s'+t'+u'=s''+t''+u''=2m^2$.
Using unitarity we can write  $\text{Im}\,\mathcal{A}_4(s,t)$ as
\begin{equation}
\begin{split}\label{ImA4}
  \text{Im}\,\mathcal{A}_4(s,t)
  ~=~ \frac{(s-4m^2)^{3/2}}{128(2\pi)^4 \sqrt{s}} \int d\Omega_5\,
  \ca^{\varphi\varphi\Phi\Phi}(t',u')\,\ca^{\varphi\varphi\Phi\Phi}(t'',u'')^*\,,
\end{split}
\end{equation}
where $\Omega_5$ is the standard measure on $S^4$; it integrates over the spatial direction of the $q_i$ in the center-of-mass frame. A factor of 1/2 was included to account for identical final states. 
The two subamplitudes $\ca^{\varphi\varphi\Phi\Phi}$ in the bubble diagram of Fig.~\ref{figunitarity} are simply the tree-level amplitudes
\be
\ca^{\varphi\varphi\Phi\Phi}(t,u)  ~=~
\frac{1}{f^4}
\left[\frac{4m^4}{t-m^2} + \frac{4m^4}{u-m^2}\right]\,.
\ee
It is straightforward to assemble the integrand of the dispersion relation~(\ref{deltaadisp}),

\begin{equation}\label{Ds}
    {\cal D}(s)~\equiv~\frac{2 f^8}{9\pi}\biggl[
    \frac{3}{s^4}
    \text{Im}\,\mathcal{A}(s,0)-\frac{2}{s^3}\frac{\partial}{\partial t}\text{Im}\,\mathcal{A}(s,0)\biggr]\,,
\end{equation}
and find after numerical computation
\begin{equation}
          \Delta a~=~\int_{4m^2}^\infty \!\!\!ds\, {\cal D}(s)~=~\frac{1}{(4 \pi)^3 2^4 3^4 7}
~=~a_\text{sc}          
          \,.
\end{equation}
This is precisely the expected result  \reef{asc}; thus  it  
provides a check of our dispersion relation \reef{deltaadisp}. 
However, it is instructive to examine the integrand ${\cal D}(s)$ more closely. It is plotted in Fig.~\ref{figDs}. We note that ${\cal D}(s)$ is positive at small $s$:  this makes sense since we expect the $s$-wave to dominate at low energies.  In terms of the partial wave expansion,   the $s$-wave contributes only to the first term of 
${\cal D}(s)$. However, the integrand actually turns \emph{negative} for large $s$ ($ \gg m^2$). The second term in~(\ref{Ds}) thus eventually dominates the first term. However, the large $s$ 
contributions are strongly suppressed and the overall integral is positive. 
This suggests the following interpretation of the positivity of $\Delta a$: for the free massive boson, the integral is dominated by the large threshold contribution of the $s$-wave.

Since the integrand of the dispersion relation \reef{deltaadisp} is not positive definite, it is not obvious how one might use it to prove an $a$-theorem in $d=6$. We may further note that even though the dispersion relation \reef{deltaadisp} determines $\Delta{a}$ correctly, we cannot generalize it to an $a$-function as was done in 4d, see (3.7) of \cite{zohar}. 
The reason simply is that the change of sign in the integrand means that this putative $a$-function would not be a monotonic function of the interpolating scale $\mu$ with $4m^2 \le \mu < \infty$. The example of the free massive scalar  illustrates   this.

\compactsubsection{Spontaneously broken conformal symmetry}
We focused on the case of softly broken conformal symmetry, so let us now consider spontaneous breaking. In that case, the $O(p^6)$ matrix element is given in \reef{AO4p6}, so \reef{loweng} is replaced by 
\begin{equation}\lab{loweng2}
  \ca_4(s,t) ~=~\frac{b}{2f^8}(s^2+t^2+u^2)
  +\bigg[
    \frac{3}{2}\Delta{a} - \frac{b^2}{f^4}
  \bigg]
  \frac{3 }{f^8}\, stu
~+~O(p^8)\,.
\end{equation}
The approach above would yield a dispersion relation \reef{deltaadisp} for $\Delta{a} - \tfrac{2b^2}{3f^4}$, not just $\Delta{a}$. Some comments are in order:
\begin{enumerate}
\item If the RHS of the dispersion relation \reef{deltaadisp} were always non-negative it would give a stronger than needed constraint: $\Delta{a} - \tfrac{2b^2}{3f^4} \ge 0$.
\item It turns out (and we will prove it in Sec.~\ref{s:DBI6d}) that the  
4-point dilaton amplitude actually \emph{vanishes} at $O(p^6)$ on the Coulomb branch of the 6d (2,0) theory; \ie $\Delta{a}=\tfrac{2b^2}{3f^4}$. Thus it will \emph{not} be possible to prove a  general result 
that the coefficient of the $s t u$-term in the 4-dilaton amplitude is strictly positive. This is weaker than the 4d analogue.
\item We should ask if it even makes sense to require that the coefficient of the $s t u$-term in the 4-dilaton amplitude is non-negative. If this were always true, then that places an upper bound on $b^2/f^4$ in terms of $\Delta{a}$. 
This is  different from the lower bound placed on coefficients of higher-derivative operators, such as those discussed in \cite{nimaetal}
\end{enumerate}

 It is curious to ask  whether there is a dispersive  sum rule for $b$, the coefficient of $s^2 +t^2 +u^2$, in 6d. This would imply $b>0$.  This appears to be straightforward, but
 it is necessary to consider the convergence of the dispersion integral $\int ds\,\tfrac{{\rm Im}\,\mathcal{A}(s,0)}{s^3}$ at large $s$.  
 In $d=6$, \eqref{uvlim} is replaced by 
${\rm Im}\,A(s,0) \sim   s^{3-\eps}$ 
with $\eps = 6 -\Delta$,  where $\Delta$ is the scale dimension of the controlling relevant operator. $\eps > 1$ is 
needed to validate the sum rule, while a generic relevant operator only gives  $\eps>0$. Nonetheless, in the two cases we study explicitly in this paper, the relevant operator has $\D =4$, so $\eps=2$. Indeed we find in both cases $b>0$; see Secs.~\ref{s:massive} and \ref{s:DBI6d}. We note that the causality-based argument of \cite{nimaetal} appears to yield $b>0$ without restrictions on $\eps$.

\compactsubsection{Sum rules for the 6-point dilaton amplitude?}
The anomaly flow $\D a$ is also captured by the low-energy behavior of the 6-dilaton amplitude.  We now discuss an attempt to extract a sum rule that establishes positivity.  There are difficulties due to the more complicated kinematics, analyticity and unitarity properties of six-point functions. We will only scratch the surface of the subject here. 

Focusing first on the scenario with {\bf \em explicitly broken} conformal symmetry, it follows from \reef{Amp-p4th} and \reef{AO6p6} that the low-energy expansion is 
\bea
\mathcal{A}_6 &=& \frac{3b}{f^{12}}  \bigg[\sum_{1\le i<j\le 6} s_{ij}^2\bigg]
~+~ 
 \frac{3\Delta{a}}{f^{12}}
  \bigg[
  ~3 \sum_{1\le i<j\le 6} s_{ij}^3
  ~~~+~ \sum_{1\le i<j<k\le 6} s_{ijk}^3\,
  \bigg] ~+~ O(p^8)\,.
\eea
A natural generalization of the forward limit to six particles is 
\bea
   p_4 \to - p_1 \,,~~~~~~~
   p_5 \to - p_2 \,,~~~~~~~
   p_6 \to - p_3 \,.
  \labell{6ptforward}
\eea
In this limit, the 6-point amplitude is characterized by the three kinematic invariants $s_{12}$, $s_{23}$ and $s_{13}$; we find
\bea
\mathcal{A}_6 &=& \frac{12b}{f^{12}}  
\Big[ s_{12}^2 + s_{13}^2 + s_{23}^2 \Big]
~+~ 
 \frac{144\Delta{a}}{f^{12}}
  s_{12} s_{13} s_{23}
   ~+~ O(p^8)\,.
\eea
Now choose a frame in which the spatial momenta $k_j$, $j=1,2,3$ are coplanar, at angles $\theta_j= 2\pi j/3$ and with magnitude $|k_j|\equiv \omega_j$. It is convenient to take $\omega_1=\omega_2= \omega$ to be fixed and consider a dispersion relation in $\omega_3$. The kinematic invariants are then
\begin{equation}
    s_{12}=3\omega^2\,,\qquad s_{23}=s_{13}=3\omega\omega_3\quad\Rightarrow\quad s_{123}=3\omega(\omega+2\omega_3)\,.
\end{equation}
The $O(p^6)$ contribution to the 6-point dilaton amplitude then takes the form
\begin{equation}
     \mathcal{A}_6(\omega_3,\omega)~=~ 
     \frac{2^2 3^3 \, b}{f^{12}}  \, \omega^2 (\omega^2 + 2 \omega_3^2)
     ~+~
     \frac{2^{4}3^5}{f^{12}} \Delta a\,\omega^4\omega_3^2
     ~+~ O(p^8)\,.
\end{equation}
We therefore attempt to compute $\Delta a$ from the dispersion relation
\begin{equation}\label{Deltaa6pt}
\Delta a ~\equiv~
    \frac{f^{12}}{2^{7}3^6}\frac{\partial^4}{\partial \omega^4}\oint\frac{d\omega_3}{2\pi i\, \omega_3^3}\, \mathcal{A}_6(\omega_3,\omega)\,.
\end{equation}

In the case of the free massive scalar, only two inequivalent branch cuts contribute to this dispersion relation when $0<\omega\ll m$: the $s_{123}$-cut and the $(s_{13}\!=\!s_{23})$-cut. The contribution to $\Delta a$ in~(\ref{Deltaa6pt})
from the $s_{123}$-cut is not manifestly positive definite, despite the positive factor of  $|A(1,2,3,q_1,q_2)|^2$ that it contains by unitarity. 
As in the 4-point example above, the reason lies in the appearance of derivatives w.r.t.~$\o$
in~(\ref{Deltaa6pt}), which spoil manifest positivity. In fact, the integrand of the dispersion relation is again not positive definite in the free massive boson example. Our choice of kinematics is therefore not suitable to define a monotonic $a$-function at all scales. 

It is also useful to consider the case of {\bf \em spontaneously broken} conformal symmetry, where the $O(p^6)$ amplitude is now given by  \reef{AO6p6}. Focusing on the polynomial terms only, one might think that the LHS of the attempted sum rule \reef{Deltaa6pt} now is $\Delta{a} - \frac{5b^2}{8f^4}$. However, among the 10 non-polynomial diagrams in the 2nd line of \reef{AO6p6}, 6 diagrams actually diverge in the forward limit \reef{6ptforward}. So this indicates that the forward limit \reef{6ptforward} is simply not the correct choice of kinematics for a sum rule for the 6-particle amplitude. A smarter choice of kinematics may remedy the problems we have outlined here.  Help in the literature (for example \cite{russians}) may be available for those with stamina.

\setcounter{equation}{0}
\section{Example of spontaneous breaking: 4d $\mathcal{N}=4$ SYM}
\label{s:DBI4d}
The prime example of a 4d theory with spontaneously broken conformal symmetry is $\mathcal{N}=4$ SYM on the Coulomb branch. We will study this system from the standpoint of its gravity dual. This is  a warmup for an analogous study of the 6d (2,0) theory in Sec.~\ref{s:DBI6d}.

To set the stage for this discussion, let us take $\mathcal{N}=4$ SYM on the Coulomb branch by turning on a VEV  such that the $SU(N+1)$ gauge group is broken to $SU(N)\times U(1)$. At energy scales much greater than the VEV, the theory is the $\mathcal{N}=4$ superconformal theory with $SU(N+1)$ gauge group; this will be our CFT$_\text{UV}$.
 In the IR, the massive multiplets decouple to leave a free $U(1)$ multiplet and a CFT$_\text{IR}$ which is $\mathcal{N}=4$ SYM theory with $SU(N)$ gauge group. The $U(1)$ multiplet contains 6 scalars; one is the Goldstone boson of the spontaneously broken conformal symmetry, while the other five  are the Goldstone bosons of the spontaneously broken global R-symmetry $SU(4)\to Sp(4)$.

The gravity dual of
this scenario is the D3-brane configuration in  which a single brane is separated from a  stack of $N$ branes. In the limit of large $N$, this can be
modeled by a D3-probe brane in the gravitational background of $N$ D3's,
\be
ds^2= f^{-1/2} \sum_i^5 dx_i^2 + f^{1/2}(dr^2 + r^2 d\Omega_5^2)\,,
\quad \qquad
f(r)= 1+ \frac{4\pi g_s N l_s^4}{r^4}\,.
\ee
In the near horizon limit, $f(r)\to \tfrac{4\pi g_s N l_s^4}{r^4}$,  and the geometry becomes AdS$_5\times S^5$ with AdS-radius
$L^4 = R^4_{S^5} = 4\pi g_s N l_s^4$.  Using $r = L^2/z$ we get the Poincare patch metric,
\be \lab{ppatch-intro}
ds^2 = \frac{L^2}{z^2} \Big( \eta_{\m\n}dx^\m dx^\n + dz^2\Big)
    + L^2 \, d\Omega_5^2\,.
\ee
The dynamics of a D3-probe brane in the background \reef{ppatch-intro}
is governed by the DBI action:\footnote{Since we are only interested in the dilaton, we  ignore  motion on $S^5$.}
\be
  S_\text{D3}
  ~=~ -T_\text{D3} \int d^4x ~\frac{L^4}{z^4}
  \left(\sqrt{1+ (\partial z)^2}-1 \right)
  ~=~
  T_\text{D3} L^4  \int d^4x
  \left(
  -\frac{1}{2} \frac{(\partial z)^2}{z^4}
  + \frac{1}{8} \frac{(\partial z)^4}{z^4} + O(\pa^6) \right)
  \,.
  \label{Sd3}
\ee
The D3-brane tension is $T_\text{D3}=1/((2\pi)^3g_s l_s^4)$.
The term following the square root is the pullback of the 5-form flux, and the fact that it cancels the zero-derivative term in the expansion of the square root is the well known no-force condition.

The DBI action captures the physics of the dilaton because the $U(1)$ supermultiplet with the dilaton lives on the D3-brane.  Indeed we will find below that the DBI field $z(x)$, which describes the radial motion of the probe brane,  is related by a change of field variable to the dilaton $\t(x)$.
When the change of variable  is made, the DBI action becomes the dilaton action through 4th order in a derivative expansion. First, however, we do something simpler;  we compute the low energy scattering amplitude directly from the DBI action and compare it with
the results found from the dilaton effective action in Sec.~\ref{s:varphi}.
We will learn that the coefficient   $T_\text{D3}L^4$ 
of the DBI action determines the anomaly flow
 $\Delta{a}=a_\text{UV} - a_\text{IR}$, at leading order in large $N$.

\subsection{Matrix elements of the radial mode on the D3-brane DBI action}
To calculate scattering processes, we perform a field transformation from $z$ to $\phi$ via $z = L \,e^\t$ and $e^{-\t} = 1 -\frac{\phi}{f}$. This gives
\bea
  S_\text{D3}
  ~=~
  \int d^4x
  \left(
 -\frac{1}{2}  \frac{T_\text{D3} L^2}{f^2} (\pa \phi)^2
  + \frac{1}{8}  \frac{T_\text{D3} L^4}{f^4} (\pa \phi)^4
  +\dots
  \right) \,.
  \label{N4SYM-phi}
\eea
The kinetic term is canonically normalized if
\be
\label{f4d}
f^2 =T_\text{D3} L^2\,.
\ee
The 4-point matrix element is then
\bea
  A_4^{O(p^4)} ~=~\frac{T_\text{D3} L^4}{4f^4} \big(s^2 + t^2 + u^2\big) \,.
\eea
Comparing this with \reef{4ptle}, we have
\bea
  \label{Da-N4SYM}
  \Delta a
  ~=~ \frac{T_\text{D3} L^4}{16}
  ~=~ \frac{N}{32 \pi^2} \,.
\eea

The  $a$-charge of $\mathcal{N}=4$ SYM (in the large $N$ limit)
is $a= N^2/(64\pi^2)$  \cite{HenSken} , and therefore
\be
\Delta a = a_\text{UV}- a_\text{IR} = \frac{N^2- (N-1)^2}{64\pi^2} = \frac{N}{32\pi^2} +
O(N^0)\,.
\ee
This matches precisely the result \reef{Da-N4SYM} of the probe brane action to leading order in large $N$.

The field redefinition above \reef{N4SYM-phi} indicates a relation between the radial mode $z$ and the dilaton $\tau$. However, if $z = L \,e^\t$ is plugged into the DBI action \reef{Sd3}, we do \emph{not} recover the off-shell 4d effective action \reef{dilact}. We must be more careful in identifying the dilaton, and this requires a closer examination of the symmetries of the DBI action. Since the following analysis is also useful for our 6d analysis in the Sec.~\ref{s:DBI6d}, we carry it out in general $d$-dimensions.

\subsection{Symmetries: bulk diffeos and brane Weyl transformations}
\label{s:DBIsyms}

The DBI system describes the motion of a probe D$p$-brane with world volume coordinates $\xi^a, ~ a=0,\ldots, p$ in a background spacetime with  coordinates
$X^M, ~M=0,\ldots, d$ with $d > p$.  We will take the background to be AdS$_{p+2}$  with metric
$g_{MN}(X)$. 
The DBI action involves the induced metric on the brane:
\be \lab{sdbi}
S_\text{DBI} = - T_p\int d^{p+1}\xi \sqrt{-\det\left( \frac{\pa X^M}{\pa \x^a}g_{MN}\frac{\pa X^N}{\pa \x^b}\right)}\,,
\ee
where $T_p$ is the brane tension. In this general form, the symmetries consist of independent diffeomorphisms of  the $\xi^a$ and the $X^M$.  For most purposes it is
useful
to gauge-fix the diffeos of the $\xi^a$, and we use the familiar static gauge condition: $\xi^a = X^a, ~a=0,\ldots, p$.  We choose  the Poincar\'e patch metric on AdS with radial coordinate $z$, but unspecified transverse coordinates
$X^\m \to x^\m, ~\m= 0,\ldots, p$, thus the line element
\be \lab{ppatch}
ds^2 = \frac{L^2}{z^2} \Big( g_{\m\n}(x)dx^\m dx^\n + dz^2\Big)\,.
\ee
The DBI action that determines $z(x^\m)$  then becomes:
\be \lab{sdbi2}
S_\text{DBI} = -T_p\int d^{p+1}x \frac{\sqrt{-g}}{z^{p+1}} \sqrt{1 + g^{\m\n}\pa_\m z \pa_\n z}\,.
\ee
Ultimately we will take the standard Poincar\'e patch metric with $g_{\m\n}(x)\to \eta_{\m\n},$  but we wish to study a symmetry of the general form \reef{sdbi2} which will turn out to be closely related to Weyl invariance.

For this purpose we note that the form \reef{ppatch} of the AdS line element is preserved by the following variant of the well known  Penrose-Brown-Henneaux-Fefferman-Graham diffeomorphism
\be
\d_\s x^\m = -\frac12 z^2 g^{\m\n}\pa_\n \s\,, \qquad\quad
\d_\s z = \s\, z\,.
\ee
The transverse metric varies as
\be \lab{varhe1}
\d g_{\m\n} = 2\s\, g_{\m\n} +\frac12 z^2(D_\m  \pa_\n +D_\n \pa_\m)\s \,,
\ee
which consists of a Weyl transformation plus diffeomorphism.
This is not quite a symmetry of the brane action \reef{sdbi2} because the static gauge condition is violated.  Instead, one can verify that \reef{sdbi2} is invariant under the infinitesimal variation \reef{varhe1} combined with
\be  \lab{varhe2}
\d z(x)
= \s(x) z(x) +  \frac12 (z(x))^2 g^{\m\n}(x) \pa_\m\s(x) \pa_\n z(x)\,,
\ee
which now includes the pullback of the diffeomorphism $\d_\s x^\m$ to the brane. This symmetry of $S_\text{DBI}$ in static gauge is the extension to a general
brane metric $g_{\m\n}(x)$ of the argument of  Sec.\ 2 of \cite{malda97} that the DBI action with Cartesian Poincar\'e patch coordinates (\ie $g_{\m\n} \to \eta_{\m\n}$)
is invariant under global special conformal transformations.

There is a general AdS/CFT relation between radial evolution in AdS spacetime and RG flow
in the dual field theory which suggests that there should be a change of radial coordinate which relates $z(x)$ to the dilaton field $\t(x)$.
As the first step to implement this idea, we pose the tentative  relation  $z(x) = L\,e^{\t(x)}$.  It is clear that the Weyl transform $\d\t = \s$  produces  the first term of \reef{varhe2}.  However, this is not enough; we must correct the tentative relation by adding terms involving derivatives of $\t$ to make  $\d\t = \s$ produce \reef{varhe2}  exactly.  When this correction is implemented to the appropriate order in derivatives, the DBI action, expressed in terms of $\t$,  {\it becomes} the dilaton effective action!

\subsection{From $z(x)$ to $\tau(x)$}
\label{s:ztau}
We write an ansatz for this relation between the radial mode $z$ and the dilaton $\t$ as
\bea
  \label{z-tau}
  z = L \,e^\tau
  + L^3 \,e^{3\tau}
  \Big[ \alpha_1 \, \Box \tau
  + \alpha_2 \, (\partial\tau)^2
  + \alpha_3 \, R  \Big] + O(\pa^4) \, .
\eea
Our aim is to find relations among the dimensionless constants $\alpha_i$ such that a Weyl variation of the RHS of \reef{z-tau} matches
$ \d z(x) $ of \reef{varhe2}.   We work consistently to second order in derivatives in this section and consider higher order corrections in Sec.~\ref{s:DBI6d}. Keeping only terms up to second order, \reef{varhe2} can be written as\footnote{Note that the presence of the $z^2$-term in \reef{varhe2} means that the derivative-expansion  \reef{z-tau} does not truncate at finite order and the change of variables from $z$ to $\tau$ is therefore non-local.}
\bea
  \label{bulk-dz2}
  \delta z
  ~=~ \sigma\,z
   +\tfrac{1}{2} L^3 \, e^{3\tau} \pa_\m \tau\, \pa^\m \sigma
   +  O(\pa^4)
  \,.
\eea

To second order in derivatives, the variation of the RHS of \reef{z-tau}
may be calculated using $\d\t=\s$ and the  Weyl part of \reef{varhe1}.  To calculate it we note that
\bea
\nonumber
  \d \big( e^{3\tau} \Box \tau  \big)
  &=& e^{3\tau}
  \big[
    \,\s \, \Box\tau + \Box\sigma + (d-2) \pa\tau\cdot\pa\s
  \big]
    \label{1st-var}
  \big]\,,~~~~
  \\[3mm]
  \d\big( e^{3\tau} (\pa\t)^2  \big)
  &=& e^{3\tau}
  \big[
    \s \, (\pa\t)^2 + 2 \pa\tau\cdot\pa\s
  \big]\,,
  \\[3mm] \nonumber
  \d \big( e^{3\tau} R  \big)
  &=& e^{3\tau}
  \big[
    \s \,R - 2(d-1) \Box\s
   \big] \,.
\eea
Inserting these results  in \reef{z-tau}, we find
\bea
  \big( \delta_\sigma z \big)_\text{brane}
  &=&
     \s \,  z
  + L^3 \,e^{3\tau}
  \Big[  \big(  \a_1(d-2) + 2 \a_2 \big) \pa \t\cdot\pa \sigma
  +\big( \a_1  -2 \a_3 (d-1)   \big) \Box \s
  \Big]
  +  O(\pa^4)\, .~~~~~~~
  \label{brane-dz2}
\eea

The two expressions \reef{bulk-dz2} and   \reef{brane-dz2} match
if  we impose the constraints
\be
  \label{a1a2a3}
  \a_3 = \frac{1}{2(d-1)} \a_1 \,,
  ~~~~~~~
  \frac{d-2}{2} \alpha_1 + \alpha_2 = \frac{1}{4} \,.
\ee
This will play a role in our analysis of both the D3-brane and the M5-brane effective actions.
In both cases, we will be working with a flat metric on the brane, so the curvature term with coefficient $\alpha_3$ in \reef{z-tau} vanishes
and all covariant derivatives become partial derivatives.
 Using the second condition of \reef{a1a2a3} we find the net result
\be
  \label{z-tau-again}
  z = L \,e^\tau
  + \frac{1}{4}L^3 \,e^{3\tau}Q
   + O(\pa^4) \, ,
  ~~~~~~\text{with}~~~~~~~
  Q=\Big[ (\pa \tau)^2 + 4 \alpha_1 \, \Big( \Box \tau
  - \frac{d-2}{2} \, (\pa \tau)^2 \Big)
  \Big]\,.
\ee
Note that when then $d$-dimensional EOM of $\tau$ is imposed, the dependence on $\a_1$ drops out completely and we are left with a much simpler expression.
We will use the result \reef{z-tau-again} in the analysis of a probe brane in the AdS$_5$ and AdS$_7$ backgrounds.

\subsection{Matching the D3-brane DBI action and $S_\text{eff}[\tau]$}
\label{s:D3action}

The D3 probe action in the AdS background \reef{ppatch-intro} was given in \reef{Sd3}. Now we use the relation \reef{z-tau-again} between $z$ and $\tau$: expanding and keeping only 4-derivative terms we find
\bea
  \frac{(\pa z)^4}{z^4} &=& (\pa \t)^4 + O(\pa^6)\,,\\[1mm]
  \nonumber
  \frac{(\pa z)^2}{z^4} &=& \frac{1}{L^2} (\pa \t)^2 e^{-2\t}
  + \frac{1}{2} (\pa \t)^2 Q + \frac{1}{2} \pa_\m \t \pa^\m Q +  O(\pa^6) \\
  &=& \frac{1}{L^2} (\pa \t)^2 e^{-2\t}
  - \frac{1}{2} \big[ \Box \t - (\pa \t)^2 \big] Q +  O(\pa^6)\,.
\eea
In the last line we used partial integration. Inserting the expressions above into \reef{Sd3}, we find
\be
  \label{SD3-B}
  S_\text{D3}
  =  \int d^4x
  \left(
  -\frac12 
  f^2 
  (\pa \t)^2 e^{-2\t}
  +
  \frac{T_\text{D3} L^4}{8}
  \Big[
    2 \big( \Box \t - (\pa \t)^2 \big) Q + (\pa \t)^4
  \Big]+ O(\pa^6) \right),
\ee
where we have identified the dilaton decay constant $f$ as in \reef{f4d}. We substitute  $Q$ from \reef{z-tau-again} and  find
\be
  \label{SD3-C}
  S_\text{D3}
  =  \int d^4x
  \left(
  -\frac12 f^2 (\pa \t)^2 e^{-2\t}
  +  \a_1 T_\text{D3} L^4  \Big( \Box \t - (\pa \t)^2 \Big)^2
  +
  \frac{T_\text{D3} L^4}{8}
    (\pa \t)^2 \Big( 2 \Box \t - (\pa \t)^2 \Big)
  + O(\pa^6) \right).
\ee
Lo and behold! Our reader will  recognize this as the off-shell dilaton effective action \reef{dilact} with $\a=0$ and $\kappa = \a_1 T_\text{D3} L^4$. Note that the  constant $\a_1$ that remained undetermined by the bulk-brane symmetries in the previous section is proportional to the coefficient of the $\hat{R}^2$-term which vanishes on-shell.

By comparison to \reef{dilact}, the coefficient of the last term of \reef{SD3-C} is $2 \Delta a$, so we have
\bea
   \Delta a = \frac{T_\text{D3} L^4}{16} = \frac{N}{32\pi^3} \,,
\eea
using $T_\text{D3}=1/((2\pi)^3g_s l_s^4)$ and $L^4=4\pi g_s N l_s^4$.
As discussed previously, this matches the expected value.

\vspace{2mm}
The DBI action is itself an effective action, and the fact that it produces the dilaton effective action correctly up to order $O(\pa^4)$ is an indication that the DBI action captures correctly the 4-derivative terms of the brane-dynamics (see for example \cite{BPT} and references therein).

\setcounter{equation}{0}
\section{Example of spontaneous breaking: 6d (2,0) theory}
\label{s:DBI6d}

At large $N$, we can study the Coulomb branch of the 6d (2,0) theory by placing an M5-brane in the background of $N$ M5-branes. The geometry is
\bea
  ds^2 = h^{-1/3} dx^2_\text{6d} + h^{2/3} (dr^2 + r^2 d\Omega_4^2) \, ,
  \quad \qquad
  h(r) = 1 + \frac{\pi N \ell_p^3}{r^3} \, .
\eea
The near-horizon limit takes $h(r) \to \pi N \ell_p^3/r^3$, and we find the near-horizon geometry AdS$_7\times S^4$.  Poincare coordinates \reef{ppatch} are obtained from $r = L^3/(4z^2)$.

The 6d DBI action is
\bea
  \nonumber
  S_\text{DBI} &=&
  - T_\text{M5} \,
  \int d^6x ~ \frac{L^6}{z^6} \Big( \sqrt{1+(\pa z)^2}  -  1\Big)\\[2mm]
  &=&
  - T_\text{M5} \, L^6 \int d^6x ~\bigg(
  \frac{1}{2}\frac{(\pa z)^2}{z^6}
  - \frac{1}{8}\frac{(\pa z)^4}{z^6}
  + \frac{1}{16}\frac{(\pa z)^6}{z^6} + O(\pa^8)
  \bigg)\,,
  \label{6dS-DBI}
\eea
where the M5-brane tension and the AdS$_7$ radius are
\bea
  T_\text{M5}  = \frac{1}{(2\pi)^5 \ell_p^6}\,,
  \hspace{1.3cm}
  L = 2R_{S^4}= 2 (\pi N)^{1/3} \ell_p\,.
  \label{TM5andL}
\eea

As in the 4d case of $\mathcal{N}=4$ SYM we compute $\Delta{a}$ in two different ways. First by matching the matrix elements of scattering of the radial mode of the DBI action to those of  Sec.~\ref{s:varphi} and then by matching the  DBI action \reef{6dS-DBI} to the 6d dilaton effective action.

\subsection{Matrix elements of the radial mode on the D3-brane DBI action}

To calculate matrix elements, we introduce $\phi$ via
$z = L \,e^\t$ and $e^{-2\t} = 1 -\frac{\phi}{f^2}$. The result is
\bea
  \nonumber
     S_\text{DBI} &=&
    \int d^6x ~\bigg(
   -\frac{1}{2} (\pa \phi)^2
   + \frac{T_\text{M5} L^4}{8}
     \Big[
       \frac{1}{16f^8} (\pa \phi)^4
       + \frac{3}{16f^{10}} \phi (\pa \phi)^4
       + \frac{3}{8f^{12}} \phi^2 (\pa \phi)^4
     \Big]
     \\[2mm]
     && \hspace{2.2cm}
     -  \frac{T_\text{M5} L^6}{2^{10} f^{12}}  (\pa \phi)^6
     + O(\phi^7,\pa^8)
   \bigg)\,.
   \label{6dDBI-phi}
\eea
From \reef{Seff6dGEN} we  identify
\bea
  \label{our-f}
   f^4 = \tfrac{1}{4}T_\text{M5} L^2\,.
\eea

The $O(p^4)$ matrix elements are found directly from the contact terms:
\bea
  \nonumber
  A_4^{O(p^4)} &=&
  \frac{T_\text{M5} L^4}{2^6 f^8}(s^2 + t^2 +u^2) \, ,\\[2mm]
  \nonumber
  A_5^{O(p^4)} &=&
  \frac{3 T_\text{M5} L^4}{2^7 f^{10}}  \sum_{1\le i<j\le 5} s_{ij}^2\, , \\[2mm]
  A_6^{O(p^4)} &=& \frac{3T_\text{M5} L^4}{2^5 f^{12}}  \sum_{1\le i<j\le 6} s_{ij}^2\, .
  \label{Amp-p4th-DBI}
\eea
These are completely consistent with the matrix elements in \reef{Amp-p4th} and identify
\bea
    b ~=~ \frac{T_\text{M5} L^4}{2^5}
    ~~~~~
     \implies
    ~~~~~
    \frac{b^2}{f^4}~=~ \frac{N^2}{2^7 \pi^3}
    \,.
    \label{theb}
\eea

Simple inspection makes it clear that the action \reef{6dDBI-phi} does not produce any 4- and 5-point matrix elements at order $O(p^6)$:
\bea
  A_4^{O(p^6)} ~=~ 0 \,,
  ~~~~~~~~~~~
  A_5^{O(p^6)} ~=~ 0 \,.
\eea
Comparing that with the matrix elements \reef{AO4p6} and \reef{AO5p6} shows that
$\big[ \tfrac{3}{2}\Delta{a} - \tfrac{b^2}{f^4}\big]$
must vanish; with the help of \reef{TM5andL} and \reef{our-f} this fixes the change in the $a$-charge to be $\Delta{a} ~=~ 2b^2/(3f^4)$ which by \reef{theb} is
\bea
  \label{6d-Delta-a}
  \Delta{a}
  ~=~ \frac{N^2}{192 \pi^3}\,.
\eea
This is in fact the correct value at large $N$. Indeed, the $a$-charge
of the 6d (2,0) theory was found by \cite{baste} to be
$a= N^3/(576\pi^3)$ for large $N$.\footnote{Corrections subleading in $N$ were studied in \cite{AT}.}
 Therefore
\be
a_\text{UV}- a_\text{IR}
~=~ \frac{N^3- (N-1)^3}{576\pi^3}
~=~ \frac{N^2}{192\pi^3} +
O(N^1).
\ee
It is worth noting that the match of the on-shell matrix elements \emph{required} that the dilaton was dynamical: the tree-level exchange diagrams were responsible for the $b^2/f^4$ terms in the matrix elements \reef{AO4p6} and \reef{AO5p6}, and without them the DBI matrix elements could not have been matched.

We have determined $\Delta a$, but for consistency it is also instructive to compute the 6-point DBI matrix element at $O(p^6)$. The DBI action \reef{6dDBI-phi} produces a local term from the $O(\pa^6)$ contact term as well as a non-local term contribution via the pole terms constructed from two $O(\pa^4)$ 4-vertices.
The contact term gives
\be
   A_6^{O(p^6)}
   ~=~
   -\frac{3 T_\text{M5} L^6}{2^9 f^{12}}
   \Big( s_{12} s_{34} s_{56} + \text{perms}\Big)
   ~=~
   -\frac{T_\text{M5} L^6}{2^{10} f^{12}} \Big(
   2 \sum_{1\le i<j\le 6} s_{ij}^3
      - \sum_{1\le i<j<k\le 6} s_{ijk}^3
   \Big)\,.
   \label{CT-DBIp6}
\ee
The value of the pole diagram constructed from two 4-vertices of the action \reef{6dDBI-phi} is
\be
   \text{pole-diag} ~=~
   \frac{T_\text{M5}^2 L^8}{2^{12} f^4}
   \frac{1}{f^{12}}
   \bigg(
      \Big(s_{12}^2+s_{13}^2+s_{23}^2 - s_{123}^2\Big)
   \times \frac{-1}{s_{123}}\times
       \Big(s_{45}^2+s_{46}^2+s_{56}^2- s_{456}^2\Big)  + \text{perms}
   \bigg) \,.
\ee
This contains both local and non-local terms. The result of adding  all permutations is
\bea
  \nonumber
   \text{pole-diag}
  &=& -\frac{T_\text{M5}^2 L^8}{2^{12} f^4}
   \frac{1}{f^{12}}
   \bigg(
      \Big(s_{12}^2+s_{13}^2+s_{23}^2\Big)
   \times \frac{1}{s_{123}}\times
       \Big(s_{45}^2+s_{46}^2+s_{56}^2\Big)  + \text{perms}
   \bigg)
   \\[2mm]
   &&
   +
   \frac{T_\text{M5}^2 L^8}{2^{12} f^4}
   \frac{1}{f^{12}}
   \bigg(
    2 \sum_{1\le i<j\le 6} s_{ij}^3
      - \frac{1}{2} \sum_{1\le i<j<k\le 6} s_{ijk}^3
     \bigg) \,.
     \label{pole-DBIp6}
\eea
With the help of \reef{TM5andL} and \reef{our-f}, we have
\bea
  T_\text{M5} L^6 = \frac{2N^2}{\pi^3}\,,
  ~~~~~~~~~~~~
  \frac{T_\text{M5}^2 L^8}{f^4} =
  4 \,T_\text{M5} L^6
  = \frac{8 N^2}{\pi^3} \,.
\eea
When these values are used above, we find that the $\sum_{1\le i<j\le 6} s_{ij}^3$-terms in \reef{CT-DBIp6} and \reef{pole-DBIp6} exactly cancel and we are left with the final result
\bea
  \nonumber
  A_6^{O(p^6)}
  &=&
   \frac{N^2}{2^{10} \pi^3}
   \frac{1}{f^{12}}
      \sum_{1\le i<j<k\le 6} s_{ijk}^3
  \\[3mm]
    &&
   ~-~
   \frac{N^2}{2^{9} \pi^3}
   \frac{1}{f^{12}}
   \Big(
      \big(s_{12}^2+s_{13}^2+s_{23}^2\big)
   \times \frac{1}{s_{123}}\times
       \big(s_{45}^2+s_{46}^2+s_{56}^2\big)  + \text{perms}
   \Big)\,.
      ~
    \label{AO6p6-DBI}
\eea

Let us compare this with the result \reef{AO6p6} from the dilaton effective action.
The coefficient of the $s_{ij}$-terms in \reef{AO6p6} is
$\big[ \tfrac{3}{2}\Delta{a} - \tfrac{b^2}{f^4}\big]$,
so by the discussion above \reef{6d-Delta-a}, it vanishes. This is consistent with the absence of $s_{ij}$-terms in \reef{AO6p6-DBI}. Next, the coefficient of
the $s_{ijk}$-terms in \reef{AO6p6} is
\bea
   \Big[
    \Delta{a}
      - \frac{5b^2}{8f^{4}}
   \Big]
   \frac{3}{f^{12}}
   =
   \Big[
    \frac{N^2}{192 \pi^3}
      - \frac{5}{8} \frac{N^2}{2^7 \pi^3}
   \Big]
   \frac{3}{f^{12}}
   =
   \frac{N^2}{2^{10} \pi^3} \,.
\eea
It matches the coefficient of the $s_{ijk}$-terms in \reef{AO6p6-DBI}!
Finally,  the coefficient of the pole term in \reef{AO6p6} is
$ \frac{b^2}{4f^{4}}\frac{1}{f^{12}}=
\frac{N^2}{2^9 \pi^3}\frac{1}{f^{12}}$. Again, this matches the pole term coefficient of \reef{AO6p6-DBI} perfectly.

\subsection{Matching the M5-brane DBI action and $S_\text{eff}[\tau]$}

We have seen that the matrix elements of the radial mode exactly match those of the dilaton.  In this section, we identify the dilaton $\tau$ from the radial mode $z$.
The strategy is simply an extension of the analysis described in
Secs.~\ref{s:DBIsyms}-\ref{s:ztau} and applied to the 4d D3-brane action in Sec.~\ref{s:D3action}.

The starting point is to extend the derivative expansion \reef{z-tau-again} of $z$ in terms
$\t$ as
\bea
  \label{z-tau-QW}
  z = L \,e^\tau
  + \frac{1}{4}L^3 \,e^{3\tau}\,Q
  + L^5 \,e^{5\tau}\,W
   + O(\pa^6) \, .
\eea
with $Q$ given in \reef{z-tau-again}
and
\bea
\nonumber
 W &=&
   \b_1\, \Box^2\tau
   + \b_2 \,(\Box \tau)^2
   + \b_3 \,(\partial\partial \tau)^2
   + \b_4 \,(\partial_\mu \tau) (\pa^\mu \Box \tau)
   + \b_5 \,(\Box \tau)(\partial \tau)^2
   \\[1mm]
     \nonumber
   && 
   + \b_6 \,(D^\mu \pa^\nu \tau \,\partial_\mu \tau\,\partial_\nu \tau)
   + \b_7 \,(\partial \tau)^4
   + \g_1 \,(\Box R)
   + \g_2\,(\Box\tau) R
   + \g_3\,(D_\m \partial_\n \tau) R^{\m\n}
   \\[1mm]
   && 
   + \g_4\,(\partial_\m \tau) \pa^\m R
   + \g_5\,(\partial \tau)^2  R
   + \g_6\,(\partial_\m \tau) (\partial_\n \tau)  R^{\mu\n}
  \Big] \,.
\eea
The  constants $\b_i$ and $\g_i$ are dimensionless. Note that the $\g_i$-terms vanish in the flat-space limit.\footnote{There is no need to include $R^2$ or $R_{\m\n}R^{\m\n}$ because we consider linearized transformations around a flat background. Note also that we did not include a the term $\partial_\m \tau D_\nu R^{\m\nu}$ because it is equivalent to the term
$\partial_\m \tau\, \pa^\m R$ by the Bianchi identity.}

Next we vary the RHS of \reef{z-tau-QW} under a Weyl transformation $\delta\tau = \sigma$ and $\delta g_{\m\n} = 2\sigma\, g_{\m\n}$ and require that it matches  $\delta z$ in \reef{varhe2}.\footnote{One can show that the second term in the metric transformation \reef{varhe1} is actually not important for our argument since it only shifts the constants $B_1$, $B_2$, and $B_3$. So we simply drop it.} This determines a set of relations among the coefficients. Six of these fix the $\g_i$ in terms of the $\beta_i$; these do not really concern us since we are interested in the flatspace limit of \reef{z-tau-QW} which is independent of the $\g_i$-terms. In addition, however, there are four more relations which we write
\bea
  \nonumber
   \a_2 &=& \tfrac{2}{5}\big( 4\b_5 - \b_6 + 4 \b_7 \big) \,,\\
  \nonumber
  \b_1 &=& \frac{1}{64} + \frac{1}{4}\b_3 -\frac{1}{2} \b_4 - \frac{1}{8} \b_6\,,\\
  \nonumber
  \b_2 &=& \frac{5}{128} - \frac{1}{4}\b_3+ \frac{1}{4}\b_4 -\frac{1}{2} \b_5
  + \frac{1}{16} \b_6 - \frac{1}{4}\b_7\,,\\
  \beta_3 &=& \beta_4 - \frac{2}{5}\beta_5   + \frac{3}{5}\b_6 - \frac{2}{5}\b_7 \,.
  \label{cond-beta}
\eea
This can be written more systematically to express $\alpha_2$ and
$\beta_{1,2,3}$ in terms of $\beta_{4,5,6,7}$, but the above form is more useful for us.

To conclude,  diffeo/Weyl symmetry does not fix the relationship uniquely between the dilaton $\tau$ and the radial mode $z$, not even in the flat space limit. We have to carry the undetermined constants along when we compare the DBI action \reef{6dS-DBI} to the dilaton effective action.
For this comparison, we need the dilaton effective action in a form that includes the Weyl-invariant 4- and 6-derivative terms discussed in sections \ref{s:S4deriv}, \ref{s:S6deriv}, and \ref{s:Seuler}. Let us write it as 
\bea
  \nonumber
  S_\text{eff} &=&
  \int d^6 x ~\sqrt{-\hat{g}}\,\bigg(
  - \frac{f^4}{10} \, \hat{R}
  +\frac{A_1}{100}\, \hat{R}^2
  +\frac{A_2}{2}\, \hat{R}_{\m\n} \hat{R}^{\m\n}  \\[1mm]
  &&
  \hspace{2.2cm}~
  + \frac{B_1}{100}\, \hat{R}^3
  + \frac{B_2}{100}\, \hat{R}\, \hat{R}_{\m\n} \hat{R}^{\m\n}
  + \frac{B_3}{100}\, \hat{R} \, \hat{\Box}\,\hat{R}\bigg)
  + \Delta a\, S_\text{Euler} \,,~~~~~~~~
  \label{Seff6d1}
\eea
with $S_\text{Euler}$ in the form given in \reef{euler6}. Explicit expressions for the Weyl invariants can be found in App.~\ref{useful}.

The next step is to use the flat-space version of \reef{z-tau-QW} for $z$ in terms of $\t$ in the DBI action \reef{6dS-DBI}. If one does this for general $\beta_i$, then one finds that coefficients $A_i$ and $B_i$ are fixed in terms of the $\beta_i$ and, in addition,  the $\beta_i$ must satisfy the first three relations of \reef{cond-beta} and  $\Delta{a}$ is fixed to take the value \reef{6d-Delta-a}. Curiously the symmetry constraints relating $z$ and $\tau$  requires the four relations
 of  \reef{cond-beta} while the match of
  the DBI and  $\tau$ effective actions  requires only three. All in all, this gives a consistent illustration of  how the dilaton effective action arises from the M5-brane dynamics.

\setcounter{equation}{0}
\section{Summary and Discussion}
\label{s:discuss}

The main goal of this paper was  
to explore the extension of the dilaton-focused  approach to the $a$-theorem of \cite{zohar, Z2} to $d=6$ dimensions.  
Toward that end, we constructed the low-energy expansion of the dilaton effective action  $S_\text{eff}[g,\t]$. 
It is striking how symmetries of $S_\text{eff}$ 
in a general background metric determine the low energy dynamics of the dilaton in flat space, which is  summarized  in \reef{Seff6dGEN}.  The trace anomaly fixes the 6-derivative terms, and, as a new feature in $d=6$,  there is a 4-dilaton term which descends from a Weyl invariant in curved space and does not vanish on-shell.

In Sec.~\ref{s:varphi}, we tabulated the on-shell matrix elements of $S_\text{eff}$ at  $O(p^4)$ and $O(p^6)$.   Their structure differs in the two cases of RG flows with explicit or spontaneous breaking of conformal symmetry.  In the spontaneously broken case,  the dilaton is a degree of freedom of the low-energy theory. The effective action generates tree diagrams in which the dilaton is exchanged on internal lines. This complicates the extraction of the anomaly flow $\Delta a$, as discussed in Sec.~\ref{s:dispersiona}. 
For explicit breaking, the dilaton is introduced as a conformal compensator and it can be regarded as an arbitrarily weakly coupled scalar. Hence one can suppress diagrams with internal dilaton lines (or alternatively regard the dilaton as a source).

The matrix elements embody the characteristic low-energy structure associated with broken conformal symmetry in $d=6$.   It is a structure which can be tested in examples of RG flows.  Such models are not easy to find, but we   have studied two models, namely the free massive scalar field as an example
of explicit breaking and the (2,0) theory  at large $N$ 
as an example of spontaneous breaking.
The low-energy structure of dilaton amplitudes is confirmed in both cases and we reproduce the  values of $\D a = a_\text{UV}-a_\text{IR}$ obtained by more conventional methods.  The (2,0) theory is treated holographically, and we show that the low-energy scattering amplitudes of the radial mode in the DBI action of a probe M5 brane in AdS$_7 \times S^4$ agree perfectly with those of the dilaton. The diagrams with internal dilaton lines 
mentioned above make crucial contributions in this example. 
Indeed, a field redefinition is shown to relate the DBI action to $S_\text{eff}[\t]$ at orders $O(\pa^2)$, $O(\pa^4)$ and $O(\pa^6)$. To our knowledge, it was not known in the literature if the DBI action would capture correctly the dynamics of a M5-brane at these orders of the derivative expansion, but our calculations indicate that it does.

It is useful to bring weakly relevant RG flows into the discussion even though we did not study them explicitly in this paper.  In \cite{Z2},  the dilatonic formulation of the $a$-theorem for weakly relevant flows  was checked by direct computation in $d=4$,  
and that treatment is easily extended to higher dimensions.   
Thus the formulation of the $a$-theorem of \cite{zohar} does extend to $d=6$ in three distinct models.  We know no counter-examples.

One may hope for a general proof of a 6d  $a$-theorem, 
and it is natural to try to extend the dispersive approach of \cite{zohar}  which is based on causality, crossing symmetry,  and unitarity.   Here we encounter a difficulty.  
In $d=6$  the anomaly term in the 4-point dilaton amplitude vanishes in the forward direction, see \reef{loweng},  so the optical theorem cannot be used directly 
to establish positive anomaly flow $\Delta a > 0$ and it does not lead to a monotonic $a$-function.  
 We do derive and check a dispersive sum rule for $\D a$, but the integrand is not positive definite.   It is possible that a general proof of the $a$-theorem in $d=6$ can be obtained by studying the 6-point amplitude. In the case of spontaneously broken RG flows, another complication in a  general approach is that the anomaly coefficient must be disentangled from effects of the Weyl-invariant  $O(p^4)$ term, 
as the matrix elements in \reef{AO4p6}-\reef{AO6p6} indicate.

We comment briefly on  the structure of the dilaton effective action and its matrix elements in higher even dimension.  Here a new feature appears. 
 For example, in $d=8$, there is a Weyl invariant counterterm of the form $(\hat R_{\mu\nu}\hat R^{\mu\nu})^2$, which
contributes to dilaton amplitudes at the same order, namely $O(p^8)$,
 as the Euler anomaly term.
Another fact is that in $d>6$ 
the anomaly does not contribute  to the 4-point dilaton amplitude
in flat space. The higher-$d$ extension of  the construction of
App.~\ref{app:Seuler} shows that the anomaly term contributes to $n$-point dilaton amplitudes only for $n \ge (d+2)/2$.    Thus one would have to face the complicated analyticity of multi-point amplitudes to establish an $a$-theorem.

As noted in the introduction, considerable support for the general $a$-theorem appears in  the AdS/CFT correspondence \cite{gubser,flow2,cthem,jtl}.
One of the advantages of investigating RG flows in a 
holographic framework is that the results are readily extended to
arbitrary dimensions. In particular, the analysis of holographic
RG flows identified a certain quantity satisfying an inequality
analogous to eqs.~\reef{introx1} and \reef{introx2} for {\it any
dimension}, that is, for both odd and even numbers of spacetime
dimensions. Since there is no trace anomaly for odd $d$, a new interpretation was
required. Ref.~\cite{cthem}
identified the relevant quantity as the coefficient of a universal
contribution to the entanglement entropy for a particular geometry in
both odd and even $d$. These holographic results then motivated a
generalized conjecture for a c-theorem for RG flows of both odd- and
even-dimensional QFT's. For even $d$, this new central charge was shown
to precisely match the coefficient of the Euler trace anomaly
\cite{cthem}.
For odd $d$, it was shown that this effective charge could also be
identified by evaluating the partition function on a $d$-dimensional
sphere \cite{casini9} and so the conjecture is connected to the newly
proposed $F$-theorem \cite{fthem,fthem2}. It would also be interesting to
apply the new insights of \cite{zohar,Z2} to this question in odd
dimensions.

\section*{Acknowledgements}

We are grateful for discussions with Nima Arkani-Hamed, Anatoly Dymarsky, Jaume Gomis, Zohar Komargodski, Sergei  Kuzenko, Juan Maldacena, John McGreevy, Guy Moore, Joe Polchinski, Silviu Pufu, Adam Schwimmer, Amit Sever, Sav Sethi, and Arkady Tseytlin.

The research
of DZF is supported by NSF grant PHY-0967299 and by the US Department of Energy through
cooperative research agreement DE-FG-0205FR41360. 
HE is supported by NSF CAREER Grant PHY-0953232, and in part by the US Department of Energy under DOE grants DE-FG02-95ER 40899. 
Research at Perimeter Institute is supported by the Government of
Canada through Industry Canada and by the Province of Ontario through
the Ministry of Research \& Innovation. RCM also acknowledges support
from an NSERC Discovery grant and funding from the Canadian Institute
for Advanced Research.
RCM also acknowledges the hospitality of the KITP as this paper was
nearing completion. Research at the KITP is supported in part by the
National Science Foundation under Grant No. NSF PHY11-25915. 
ST is grateful to the Physics Department of the University of Heidelberg
for hospitality and to the Klaus Tschira Foundation for generous support.

\appendix

\setcounter{equation}{0}


\appendix

\setcounter{equation}{0}
\section{Useful formulae}
\label{useful}

\vspace{-4mm}
\compactsubsection{Curvature}
The following relations fix our curvature conventions and are also
useful in our calculations:
\be
\begin{array}{lcl}
[D_\mu,D_\nu]V_\rho\equiv R_{\mu\nu\rho}{}^\sigma  V_\sigma\,,
&&
R_{\mu\nu}=R_{\mu\rho\nu}{}^\rho\,, \\[2mm]
D^\mu R_{\mu\nu\rho\sigma}=D_\rho R_{\nu\sigma}-D_\sigma
R_{\nu\rho}\,,
&&
D^\mu R_{\mu\nu}={1\over2}\partial_\nu R\,,\\[2mm]
R^{\mu\rho\sigma\lambda}D_\mu R_{\nu\rho\sigma\lambda}={1\over4}
\partial_\nu(R^{\mu\rho\sigma\lambda}R_{\mu\rho\sigma\lambda})\,,
&&[\Box,\partial_\mu]\t=R_{\mu}{}^{\nu}\partial_\nu\t\,.
\end{array}
\ee
We also include the definition of the Weyl tensor, which depends the
spacetime dimension $d$,
 \be
 W_{\m\n\r\s}=R_{\m\n\r\s}-{2\over d-2} \left( g_{\m[\r}\,R_{\s]\n}-g_{\n[\r}\,R_{\s]\m}
 \right)+{2\over(d-2)(d-1)}\,R\,g_{\m[\r}\,g_{\s]\n}\ .
 \label{weyl}
 \ee

\compactsubsection{Weyl transformations}
It is useful to record the Weyl transforms of the quantities we will
encounter. Under the Weyl transformation \reef{weyltrf}, the metric and frame field transform as  $g_{\m\n}\to e^{2\s}\,g_{\m\n}$  and $e^a_\m \to e^\s e^a_\m$. 
(The frame field formulation will be useful in the next appendix.)
Thus we have
 \bea  
 \nonumber
\G^\r_{\n\s} &\to&  \G^\r_{\n\s}  + \d^\r_\n\pa_\s\s + \d^\r_\s\pa_\n\s
- g_{\n\s} \pa^\r\s \,, \\[1mm]   
\nonumber
R_{\m\n}{}^\r{}_\s  
&\to&
R_{\m\n}{}^\r{}_\s  
+\big[ -D_\m \pa^\r\s g_{\n\s} +  \d^\r_\n D_\m\pa_\s\s 
\\
&&\hspace{2.4cm} \lab{criema} 
+
\d^\r_\m\big(\pa_\s\s\pa_\n\s-(\pa\s)^2 g_{\n\s}\big) + \pa^\r\s\pa_\m\s g_{\n\s}
- ({\m \lra \n}) \big]\,,\\[1mm]     
\nonumber
R_{\m\n}{}^{ab}  &\to& R_{\m\n}{}^{ab} - 4e^{[a}_{[\m} D_{\n]}\pa^{b] }\s 
+ 4e^{[a}_{[\m}\pa_{\n]}\s \pa^{b] }\s - 2 e^a_{[\m}e^b_{\n]}(\pa\s)^2\,,\\[1mm]  
\nonumber
 R_{\m\s}&\to&  R_{\m\s}
-(d-2)D_\m\pa_\s\s - D^\r\pa_\r\s g_{\m\s}+
(d-2)(\pa_\m\s\pa_\s\s)-(d-2)(\pa\s)^2 g_{\m\s}\,,\\[1mm]  
\nonumber
 R &\to& e^{-2\s}[ R - 2(d-1) D^\r\pa_\r\s - (d-1)(d-2) (\pa\s)^2]\,.
 \eea

\compactsubsection{Energy-momentum tensor}
Our normalization for the energy-momentum tensor is
 \be
 T_{\mu\nu}=-{2\over\sqrt{-g}}{\delta W\over\delta g^{\mu\nu}}\,,
 \lab{stress}
 \ee
where $W$ is the effective field theory action.
Given this definition, it
is straightforward to see that under a (small) Weyl transformation
\reef{weyltrf}, the effective action shifts as
 \be
W(e^{2\s}g) =W(g)+\int d^dx\sqrt{-g}\,\s\ \langle T^\m{}_\m\rangle_g\
+\ O(\s^2)\ .
 \lab{stress2}
 \ee
With these conventions, the central charges in \reef{trace4} are $a=1/(360(4\pi)^2)$ and $c=1/(120(4\pi)^2)$ for a free real conformally coupled massless
scalar in $d=4$.

\compactsubsection{Euler density}
The Euler density in $d=2p$ dimensions is defined as the top form
\be
  E_{2p} = \frac{1}{2^p} \e_{a_1b_1a_2b_2\ldots a_{p}b_p} R^{a_1b_1}\wedge
R^{a_2b_2}\wedge\dots\wedge R^{a_p b_p}\,,
\ee
where $R^{ab} = R^{ab}{}_{\m\n}\,dx^\m\wedge dx^\n$ is the curvature 2-form. It can also be written as the scalar\footnote{Note that the product  
$\sqrt{-g} \, \delta^{\mu_1\,\nu_1\,\cdots\,
\mu_{p}\,\nu_{p}}_{a_1\,b_1\,\cdots\, a_{p}\,b_{p}}$ is Weyl invariant.}
 \be
E_{2p}
=
\frac{1}{2^p}\ \delta^{\mu_1\,\nu_1\,\cdots\,
\mu_{p}\,\nu_{p}}_{a_1\,b_1\,\cdots\, a_{p}\,b_{p}}\
R_{\mu_1\nu_1}{}^{a_1b_1}\,\cdots\,
R_{\mu_{p}\nu_{p}}{}^{a_{p}b_{p}}\,.
 \lab{term0}
 \ee
The anti-symmetric Kronecker product with mixed coordinate and frame indices is defined by antisymmetrization of the inverse frame field, viz.
\be \lab{kroneck}
\delta^{\nu_1\,\nu_2\,\cdots\,
\nu_{n}}_{a_1\,a_2\,\cdots\, a_{n}}\ =  n! \,
e_{[a_1}^{\n_1}\, \ldots e_{a_n]}^{\n_n}\,.
\ee
It contains $n!$ terms with coefficients $\pm 1$ and satisfies the contraction identity 
 \be
\d^{\m_1\cdots\m_{n}\m_{n+1}}_{a_1\cdots a_{n}a_{n+1}}\,e^{a_{n+1}}_{\m_{n+1}}
= (d-n)\ \d^{\m_1\cdots\m_{n}}_{a_1\cdots a_{n}}\,.
 \lab{iden}
 \ee

When $E_{2p}$ is integrated over a compact Euclidean
$2p$-dimensional manifold, the result is a topological invariant.
Evaluating  \reef{term0}  for $d=2$ simply yields $E_2=R$, while
for $d=4$, one finds
 \bea
E_4~=~R_{\mu\nu\r\s}R^{\m\n\r\s}-4R_{\m\n}R^{\m\n}+R^2
 ~=~W_{\mu\nu\r\s} W^{\m\n\r\s}-2R_{\m\n}R^{\m\n}+\frac{2}{3}R^2\,.
 \labell{e44}
 \eea
In the second line, we have used eq.~\reef{weyl} to eliminate the
Riemann tensor in favour of the Weyl tensor. 
Similarly, for 
$d=6$ (the case of the most interest here), eq.~\reef{term0} can be
written as
 \bea
E_6&=&4 W_{\m\n\r\s}W^{\m\n}{}_{\t\u}W^{\r\s\t\u}-8W_{\m\n\r\s}
W^{\m\t\r\u} W^\n{}_\t{}^\s{}_\u-6W^{\m\n\r\s} W_{\m\n\r}{}^\t R_{\s\t}
 \labell{e66}\\ 
&&\quad+\frac{6}{5}W_{\m\n\r\s}W^{\m\n\r\s}R+3 W_{\m\n\r\s} R^{\m\r}
R^{\n\s}+\frac{3}{2}R^\m{}_\n R^\n{}_\s R^\s{}_\m
-\frac{27}{20}R^\m{}_\n R^\n{}_\m R +\frac{21}{100}R^3\,.
 \nn
 \eea
With this explicit expression, we see that only the last three terms
survive if $E_6$ is evaluated on a Weyl-flat background --- and we
recover the result given in eq.~\reef{6deuler}.

\compactsubsection{Dilaton kinetic term}
Terms in the dilaton action which are invariant under Weyl transformations and diffeomorphisms can be obtained from any integrated curvature invariant by replacing $g_{\m\n} \to  \hat{g}_{\m\n}=
e^{-2\tau} g_{\m\n}$.  For example, the dilaton kinetic term used in the text
is given in $d$-dimensions by
 \bea\nonumber
-\frac{(d-2)\,f^{d-2}}{8(d-1)}\int d^dx \sqrt{-\hg}\hR &=&
-\frac{(d-2)\,f^{d-2}}{8(d-1)}\int d^dx \sqrt{-g} e^{-(d-2)\tau}\big[
R+ (d-2)(d-1)(\pa\tau)^2\big]\\
\nonumber
&=&-\frac{f^{d-2}}{2} \int d^dx
\sqrt{-g}\left[(\pa\O)^2+\frac{d-2}{4(d-1)}R\,\O^2\right]\\
 \labell{2derb}
&\longrightarrow& -\frac12\int d^dx \,(\pa\vf)^2 \,.
 \eea
In the second line we introduce the field $\O$, and in the last line  we take the flat limit and express the action in terms of the canonically normalized scalar $\vf$. These fields are related by
\be
 \Omega= \exp\!\left[-\frac{d-2}2\,\tau\right] = \left(1-\frac{\vphi}{f^{(d-2)/2}}\right)\,.
 \lab{phi2}
 \ee
It is $\vf$ that is used in  calculations of flat space scattering amplitudes of the dilaton.  Note the equivalence of the flat space equations of motion
 \be
 \Box \tau=\frac{d-2}2\,(\pa\tau)^2\,,\quad\qquad
 \Box\vphi=0\,.
 \lab{eom}
 \ee

\compactsubsection{6-derivative Weyl-invariant operators in $d=6$}
We list results for the flat space limit  of the three independent invariants
 discussed in Sec.~\ref{s:S6deriv}.    
 Multiple
partial-integrations have been  performed to simplify the expressions.
 \bea
 \sqrt{-\hat{g}}\, \hat{R}^3 &\to&
\lab{rcube}  1000 \Big( (\Box\tau)^3 - 6(\Box\tau)^2
(\partial\tau)^2 + 12\Box\tau\, (\partial\tau)^4 - 8(\partial\tau)^6
\Big) \,,\label{rcubed}
\\[2mm]
\nonumber \sqrt{-\hat{g}}\, \hat{R}\,\hat{R}_{\m\n}\,\hat{R}^{\m\n} 
&\to&
\nonumber  20 \Big(
 7 (\Box\tau)^3
 + 8 \Box\tau\, (\partial \partial \tau)^2
 - 50 (\Box\tau)^2 (\partial \tau)^2
 - 16 (\partial\tau)^2 (\partial\partial\tau)^2
\label{rricric}
\\ &&
 ~~~~~~
 + 16\, \Box\tau\, (\partial \partial \tau\, \partial \tau\, \partial\tau)
 + 120 \Box\tau\, (\partial\tau)^4
 - 80 (\partial\tau)^6
\Big) \,,\\
\nonumber \sqrt{-\hat{g}}\,
\hat{R}\,\hat{\Box}\,\hat{R} &\to& 100 \Big(
  (\Box^2\tau)\, \Box\tau
  + 6 (\Box\tau)^3
 - 8 \Box\tau\, (\partial \partial \tau)^2
 - 20 (\Box\tau)^2 (\partial \tau)^2
 + 8 (\partial \tau)^2 (\partial \partial \tau)^2\\
 \label{rboxr}
 &&
 ~~~~~~~~
  - 16 \Box\tau\, (\partial \partial \tau\,\partial \tau\,\partial \tau)
 + 24 \Box\tau\, (\partial \tau)^4
  - 16 (\partial \tau)^6
\Big) \,.
\eea
Note that the obvious contractions are left implicit in our notation
above, 
for example
$(\partial \tau)^4=(\pa^\m\t\,\pa_\m\t)^2$ and $(\partial
\partial \tau)^2= \pa^\m \pa^\n\t\,\pa_\m \pa_\n\t$.

\setcounter{equation}{0}

\section{The Euler anomaly functional for $d=6$}
\label{app:Seuler}

The purpose of this appendix is to derive a local
diffeomorphism-invariant action $S_{\rm Euler}$ of the dilaton and
metric such that a Weyl transformation \reef{weyltrf} generates the
Euler term in $T_\mu{}^{\mu} = a\, E_6$ via
 \be \lab{tauvar}
\d S_{\rm Euler}\,= \int d^6x\,\sqrt{-g} E_6\, \d \t \,,
 \ee\marginpar{}
where the Euler density $E_6$ is explicitly given in eq.~\reef{e66}.
The starting point is the action
 \be \lab{s0}
S_0 =  \int d^6x \sqrt{-g}\,\tau\, E_6\,.
 \ee
Its ``direct" $\d\t$ variation produces the desired expression
\reef{tauvar}, but there is more.  The ``indirect" variation $\d
R_{\m\n}{}^{\r\s}$ in $\d S_0$
produces unwanted terms.  Therefore we
must add additional terms to $S_0$ in order to satisfy \reef{tauvar}
exactly.

We have computed $S_{\rm Euler}$ by two related methods:
\begin{description}
\item{\bf Method 1.}
Develop the necessary additional terms $S_i$ using infinitesimal
Weyl transformations. One works iteratively to fix $S_{i+1}$ by
requiring that its direct variation cancel the indirect variation of
$S_i$. The method is reminiscent of the old Noether method in
supergravity constructions.
 \item{\bf Method 2.}
Compute the change in
$S_0$ under a finite Weyl transformation and apply the Wess-Zumino
method to determine the additional terms needed to cancel it.
\end{description}
We begin by discussing the first method in more detail because it
appears to be new.

\vspace{3mm}
\noindent {\large \bf Method 1: infinitesimal Weyl transformations}\\[1mm]
We describe the first few steps of the iterative procedure and then set the interested reader loose to complete it.
After partial integration (easily done because the derivative  of the curvature tensor vanishes by the Bianchi identity) we find the indirect variation of \reef{s0} 
\bea  
\d S_{0,\text{ind}} 
&=&  \frac{3}{2} \int d^6x \sqrt{-g} \,\d^{\m\n\r\s\l}_{abcde}
D^e\t\pa_\l\s \,R^{ab}{}_{\mu\nu} R^{cd}{}_{\r\s} 
\,.
\lab{ds0}
\eea
To cancel \reef{ds0}  we introduce the new term in the action
\be
S_1\,=\, - \frac{3}{4}\,\int d^6x \sqrt{-g}\,  \d^{\m\n\r\s\l}_{abcde} \, \,D^e\t\pa_\l\t\,R^{ab}{}_{\mu\nu} R^{cd}{}_{\r\s}\,.
\ee
The sum of its direct variations $\pa_\l\d\t =\pa_\l\s$ and  (with partial integrations) $ D^e\d\t =D^e\s$ 
gives the desired cancellation.  The indirect  (\ie $\d R^{cd}{}_{\r\s}$) variation of $S_1$ is
\be 
\lab{dS1ind}
\d S_{1,\text{ind}} = -6\,\int d^6x \sqrt{-g} \,\d^{\m\n\r\s\l}_{abcde} 
e^e_\l \,D^c\t\,(D^d\pa_\r\t) \pa_\s\s\, R^{ab}{}_{\mu\nu}\,.
\ee
Note that there is no net variation of the metric and frame fields in 
$\d( \sqrt{-g}\,e^{e\t}\d^{\m\n\r\s\l}_{abcde})$.

We need a third term in the action containing 3 $\t$ fields with 4 derivatives and one curvature tensor to cancel \reef{dS1ind}.   The form
\be\lab{S2}
S_2 = 4 \int d^6x \sqrt{-g} \,D^c\t\,(D^d\pa_\r\t) \pa_\s\t\, \d^{\m\n\r\s}_{abcd}  \,R^{ab}{}_{\mu\nu}\,
\ee
is appropriate, and we fixed the coefficient using partial integration  to show that the direct variation contains 3 identical terms.

The indirect variation of $S_2$ has two parts,  $\d S_{2,\d R}$ from varying the curvature tensor  
and  $\d S_{2,\d\G}$  from varying the connection in $D^d\pa_\r\t =e^{d\l}(\pa_\l\pa_\r\t- \G^\a_{\r\l}\pa_\a\t)$:
\bea \lab{dS2R}
\d S_{2,\d R} &=& -48 \int d^6x \sqrt{-g} \,D^c\t\,\pa_\n\t\,(D^b\pa_\r\t)  \d^{\m\n\r}_{abc} D^a \pa_\m \s\,, \\  \nonumber
 \d S_{2,\d\G} &=& - 4\int d^4x \,D^c\t\,\pa_\s\t\,e^{d\l}(\pa_\l\t\pa_\r\s\,+\,\pa_\l\s\pa_\r\t\, - g_{\l\r} \,\pa\t\cdot\pa\s)\,\d^{\m\n\r\s}_{abcd}  \,R^{ab}{}_{\mu\nu}\\   \lab{dS2Gam}
&=&  48 \int d^6x \sqrt{-g}\,
G^{\m\n}\pa_\m\t\pa_\n\t
\,\pa\t\cdot\pa\s\,,
\eea
where $G^{\m\n} = R^{\m\n}- \frac12 g^{\m\n}R$ is the Einstein tensor.
The last expression is straightforward to derive, but it requires a lot of graphite.

To cancel $\d S_{2,\d R}$,  we need a new term in the action with 4 $\t$ fields and 6 derivatives.
The term
\bea \lab{S3}
S_3  &=& 12 \int d^6x \sqrt{-g} \d^{\n\r\s}_{bcd}\,D^c\t\,(D^d\pa_\r\t) \pa_\s\t\,
 D^b \pa_\n \t\,\\
&=&-2 4\int d^6x \sqrt{-g}\big( (\Box\t)^2 - D^\m \pa^\n\t D_\m\pa_\n\tau - \frac12  R^{\m\n} \,\pa_\m\t\pa_\n\t\big)(\pa\t)^2 
  \lab{S3bis}
 \eea
 has the right structure.  
 The second form follows after (somewhat  tricky) partial integrations.  The form  of the variation \reef{dS2Gam} suggests another term in  the action,
\be \lab{S4}
S_{4} = -12 \int d^6x \sqrt{-g}\,G^{\m\n}\pa_\m\t\pa_\n\t (\pa_\r\t\pa^\r\t)\,.
\ee
It requires some creative algebra, calculus, and  differential geometry (the Ricci identity) to show that the direct variation of \reef{S3} and \reef{S4}, with the coefficients given,  cancels all terms encountered in processing \reef{dS2R} and \reef{dS2Gam}.

The next steps, left for the reader, are to compute the indirect variations
of \reef{S3} and \reef{S4} and to show that the entire process terminates
(Hallelujah!) after two more terms are added to the action:
 \bea \lab{S5}
S_5 &=& 36 \int d^6x \sqrt{-g} \,\Box\t\, (\pa\t)^4\,,\\
S_6 &=& - 24 \int d^6x \sqrt{-g} \,(\pa\t)^6\,.
\eea

The full action is the sum of the seven terms $S_0,\dots,S_6$.  In flat spacetime it reduces to the dilation action
\be \lab{Sdil}
S_\text{anom}
= \int d^6x \big[-24 (\Box\t)^4 +24 (\pa \pa\t)^2 (\pa\t)^2 +\, 36\, \Box\t\, (\pa\t)^4 \,-24\,(\pa\t)^6
\big]\,.
\ee
It agrees with the flat space limit of \reef{dilaction} below.

\vspace{3mm}
\noindent {\large \bf Method 2: Wess-Zumino trick}\\[1mm]
If one couples a CFT to an external metric $g$ and integrates out the
CFT, one obtains a non-local effective action $W(g)$, which is the
generating functional for correlators of the energy-momentum tensor.
The presence of an anomaly ${\cal A}(g)$ manifests itself as the
non-invariance of $W(g)$ under Weyl-transformations. Using
\reef{stress2}, we have
\begin{equation}
\delta_\sigma W(g)=\int d^d x\sqrt{-g}\, \sigma\, \langle T^\mu{}_\mu\rangle_g
\equiv\int d^{d}x \sqrt{-g}\, \sigma\, {\cal A}(g)\,.
\end{equation}
${\cal A}$ is a local scalar expression composed of the curvature and covariant derivatives.
${\cal A}$ vanishes in odd dimensions

Wess and Zumino \cite{WZpaper} developed a method to integrate the anomaly. They
introduced a scalar field (Goldstone boson) for each anomalous symmetry
generator and the resulting effective action is the one relevant for
the situation where the (anomalous) symmetry is spontaneously broken.
For the Weyl anomaly the Goldstone boson is the dilaton $\tau$ and the
Wess-Zumino procedure leads to the dilaton effective action in the
background of the external metric (for a derivation see \cite{spon}) 
\begin{equation}
S(\tau,g)=\int_{0}^{1} dt \int d^d x\sqrt{-g}\,\tau\, {\cal A}\big(e^{-t\tau}g\big)
\end{equation}
The integration of the Weyl-invariant terms in ${\cal A}$ are trivial while the integration
of $E_d$ requires some work. We are interested in $d=6$ for which we find
\bea
&&S_{\rm Euler}(\tau,g)=-a\int d^6 x\sqrt{-g}\Big(\tau E_6
+12 R^\mu{}_{\rho\sigma\lambda}R^{\nu\rho\sigma\lambda}\partial_\mu\tau\,\partial_\nu\tau
-3 R^{\mu\nu\rho\sigma}R_{\mu\nu\rho\sigma}(\partial\tau)^2\nonumber\\
\noalign{\vskip.1cm}
&&\quad-24 R^{\mu\rho\nu\sigma}R_{\rho\sigma}\,\partial_\mu\tau\,\partial_\nu\tau
+12 R^{\mu\nu}R_{\mu\nu}(\partial\tau)^2
-24 R^{\mu\rho}R^\nu{}_\rho\, \partial_\mu\tau\,\partial_\nu\tau
+12 R R^{\mu\nu}\partial_\mu\tau\,\partial_\nu\tau\nonumber\\
\noalign{\vskip.1cm}
&&\quad-3 R^2(\partial\tau)^2
-16 R^{\mu\nu\rho\sigma}D_\mu \partial_\rho\tau\,\partial_\nu\tau\,\partial_\sigma\tau
+16 R^{\mu\nu}D_\mu \partial_\nu\tau\,(\partial\tau)^2\nonumber\\
\noalign{\vskip.1cm}
&&\quad
-32R^{\mu\nu}\partial_\mu\tau\,\pa^\rho\tau\,D_\rho \partial_\nu\tau
+8R\, \partial_\mu\tau\,\partial_\nu\tau\,D^\mu \pa^\nu\tau
-8 R\, (\partial\tau)^2\,\Box\tau \label{dilaction} \\ 
\noalign{\vskip.1cm}
&&\quad+16 R^{\mu\nu}\partial_\mu\tau\,\partial_\nu\tau\,\Box\tau
+6R\,(\partial\tau)^4 \nonumber\\
\noalign{\vskip.1cm}
&&\quad+24(\partial\tau)^2 (D^\mu \pa^\nu\tau)(D_\mu \partial_\nu\tau)
-24(\partial\tau)^2(\Box\tau)^2
+36\Box\tau(\partial\tau)^4
-24(\partial\tau)^6\Big)\,.
\nonumber
\eea
One immediately sees that it reduces to \eqref{Sdil} in flat space where only the last line survives.
To arrive at  \reef{dilaction} we have used the transformation rule of the
Riemann tensor under a Weyl transformation and various integrations by parts so as to
end up with an expression without any non-differentiated $\tau$.
Note that the WZ procedure guarantees a priori that the dilaton action contains only a finite
number of terms.

\compactsubsection{Paneitz operators and $Q$-curvatures}
There is  a third approach \cite{none} to construction of 
$S_{\rm Euler}$ using the mathematical formalism of Paneitz operators \cite{Graham1} and $Q$-curvatures \cite{Branson}. The utility of this
approach is that it makes clear that for any (even) spacetime
dimension, the anomalous action can be reduced to the form $\int d^dx\,
\tau \,\Box^{d/2} \tau$ in flat space, as found in \reef{dilact1} and
\reef{euler6-new} for $d=4$ and 6, respectively. This facilitated the  computation of the matrix elements in Sec.~\ref{s:varphi}. Writing the interactions in this way is a general feature
of the dilaton action because of the existence of the Panietz operators
$P_k$, a family of conformally invariant generalizations of the
d'Alembertian \cite{Graham1,Graham}. The first of these,
 \be
P_1 =
 -\Box +
\frac{(d-2)}{4(d-1)} R\ \ \longrightarrow -\Box\,,
 \ee
is simply the wave operator of a conformally coupled scalar. Similarly
for higher $k$,\footnote{For odd $d$, $k$ can be any odd integer while
for even $d$, $k \le d/2$.} $P_k$ is constructed in terms of covariant
derivatives and curvatures, however, by design in flat space, $P_k \to
(-\Box)^k$. Under conformal transformations \reef{weyltrf}, they
transform simply with
 \be
\lab{pk}
 P_k
(e^{2\sigma}g) = e^{-(d/2 + k)\sigma}\,P_k(g) \,e^{(d/2 - k)\sigma}\,.
 \ee
Just as the dilaton kinetic term \reef{2derb} takes the form of the 
conformally coupled scalar, which involves $P_1$, the higher order
Weyl-invariants
 can be written using $P_k$, \eg
 \be
\int d^dx\sqrt{-g}\ \Omega^\alpha\, P_k\, \Omega^\alpha\,,
 \lab{boo2}
 \ee
where $\alpha=(d-2k)/(d-2)$. This discussion can be extended to include
anomalous Euler term but the latter requires introducing the $Q$-curvature \cite{Branson}.
They are scalars which are built from curvatures and covariant derivatives
such that  
\begin{equation}
e^{d\sigma}Q_d(e^{2\sigma}g)=Q_d(g)+P_{d/2}(g)\,\sigma\,,
\end{equation}
\ie the inhomogeneous term is linear in $\sigma$. Observe that 
$Q_2={1\over 2} R$.  


\setcounter{equation}{0}
\section{2d CFT: the D1/D5-brane system}
\label{app:DBI2d}
The D1/D5 system is
one of the best known example of AdS$_3/$CFT$_2$. 
We consider $N_5$ D5 branes wrapped on $S^1\times M_4$, where $M_4$ is $T^4$ or $K_3$, and $N_1$ D1 branes wrapped on the same $S^1$.
The ten dimensional string frame geometry
is \cite{kraus}
\be \label{D1D5bg}
ds^2 = (Z_1Z_5)^{-1/2}(- dt^2+ dx^2) + (Z_1Z_5)^{1/2}(dr^2 + r^2 d\Omega_3^2) + (Z_1/Z_5)^{1/2} ds^2_{M_4},
\ee
where $V_4$ is the volume of $M_4$ in the asymptotically flat region and
\be
Z_{1,5} = 1+ \frac{Q_{1,5}}{r^2}\,, ~~~~~~~
Q_1= \frac{(2\pi)^4 g_s N_1 l_s^6}{V_4}\,,  ~~~~~~~
Q_5 = g_s N_5 l_s^2\,.
\ee
In the near horizon limit, we drop the 1 in the harmonic functions $Z_{1,5}$, and the metric reduces to that
of AdS$_3\times S^3 \times M_4$,
\be
ds^2 = \frac{L^2}{z^2} \big(- dt^2  + dx^2 + dz^2\big) +  L^2\, d\Omega_3^2 + \sqrt{Q_1/Q_5}\,ds^2_{M_4} \,.
\ee
Here we have introduced $z = L^2/r$ and the AdS length
$L^2 = R^2_{S^3}  = \sqrt{Q_1Q_5}$. The D1/D5 solution has a dilaton, $e^{-2\Phi} = Z_5/Z_1$, which in the near-horizon limit becomes a  constant, $e^{-\Phi} = \sqrt{Q_5/Q_1}$.

To study the system on the Coulomb branch for large $N_{1,5}$, we place a single D1 or D5 brane in the background geometry of the stack of $N_{1,5}$ D1-D5 branes.
Ignoring motion in $S^3$ or $M_4$, the respective DBI actions are given by
\be
S_\text{D1,D5}= -T_\text{D1,D5}\, \sigma_{1,5} \int d^2x \, e^{-\Phi} \,\frac{L^2}{z^2}
\Big( \sqrt{1+ (\partial z)^2} -1 \Big),
\ee
where $\sigma_1 = 1$ while $\sigma_5 = \frac{Q_1}{Q_5}V_4$
accounts for volume of the D5 brane wrapping $M_4$.
Using $z = L \,e^\tau$ (as in \reef{z-tau}) and the brane tensions
\be
T_\text{D1} = \frac{1}{2\pi \,g_s\, l_s^2}\,,
~~~~~~~~
T_\text{D5} = \frac{1}{(2\pi)^5 \,g_s  \,l_s^6}\,,
\ee
the derivative expansion of the DBI action gives
\be
S_\text{D1,D5}
=
- \frac{N_{5,1}}{4\pi} \int d^2x \, \Big[ (\partial\tau)^2 + O(\pa^4) \Big].
\ee
Comparing with the expected form of the dilaton effective action\cite{Z2}
\be
S_\text{eff} = -\frac{c_\text{UV}-c_\text{IR}}{24\pi} \int d^2x\, (\partial\tau)^2
+ \dots,
\ee
we can read off the change in the central charge:
\be
\Delta c_{[\Delta N_1=1]} = 6N_5 \,,\qquad ~~
\Delta c_{[\Delta N_5=1]} = 6N_1\,. \label{D1D5charge}
\ee
This is in complete agreement with the CFT expectation since
$c= 6N_1N_5$.


\end{document}